\documentclass[traditabstract]{aa} 
\usepackage{natbib}
\usepackage{graphicx}
\usepackage{fixltx2e}
\usepackage{mathptmx}
\usepackage{txfonts}
\def\asca{{\em ASCA}}
\def\sax{{\em BeppoSAX}}
\def\xmm{{\em XMM-Newton}}
\def\c{{\em Chandra}}

\def\swift{{\sl Swift}}
\def\hst{{\sl HST}}

\def\integral{{\sl INTEGRAL}}

\def\rxte{{\sl RXTE}}

\def\rosat{{\em ROSAT}}

\def\spitzer{{\em Spitzer}}
\def\akari{{\em Akari}}
\def\herschel{{\em Herschel}}
\def\suzaku{{\em Suzaku}}
\def\iso{{\em ISO}}
\def\iras{{\em IRAS}}

\def\vlt{{VLT}}

\def\p{$\pm$}
\def\ltsim{\mathrel{\hbox{\rlap{\hbox{\lower4pt\hbox{$\sim$}}}\hbox{$<$}}}}
\def\gtsim{\mathrel{\hbox{\rlap{\hbox{\lower4pt\hbox{$\sim$}}}\hbox{$>$}}}}

\def\lsun{$L_{\sun}$} 
\def\micron{$\mu$m}

\def\aap{A\&A}

\def\apjs{ApJS}

\def\apjl{ApJL}

\def\lognh{log$N_{\rm H}$}
\def\logl{log$L$}
\def\nh{$N_{\rm H}$}

\def\xspec{{\sc xspec}}
\def\pexrav{{\sc pexrav}}

\def\oiii{[O{\sc iii}]}

\def\l{$\lambda$}

\def\neii{[Ne{\sc ii}]}

\def\lbol{$L_{\rm Bol}$}

\def\lx{$L_{\rm X}$}
\def\lmir{$L_{\rm MIR}$}

\def\lir{$L_{\rm IR}$}
\def\av{$A_{\rm V}$}
\def\rsub{$r_{\rm sub}$}
\def\ic3639{IC~3639}
\def\eso138{ESO~138--G001}
\def\ngc3281{NGC~3281}
\def\swiftja{SWIFT~J0138.6--4001}
\def\swiftj0601{SWIFT~J0601.9--8636}
\def\alphaox{$\alpha_{\rm OX}$}
\def\alphaix{$\alpha_{\rm IX}$}
\def\fbol{f$_{\rm Bol}$}
\def\fbolmir{f$_{\rm Bol}^{\rm MIR}$}
\def\fbolx{f$_{\rm Bol}^{\rm X}$}

\begin{document}
   \title{Resolving the mid-infrared cores of local Seyferts\thanks{Based on observations carried out in ESO programme 080.B-0860.}}

   \author{P. Gandhi
          \inst{1}
          \and
          H. Horst\inst{2,3,4}
          \and
          A. Smette\inst{5}
          \and
	  S. H\"{o}nig\inst{4}
	  \and
          A. Comastri\inst{6}
          \and
          R. Gilli\inst{6}
          \and
          C. Vignali\inst{7}
          \and
          W. Duschl\inst{2}
          }

   \institute{RIKEN Cosmic Radiation Lab, 2-1 Hirosawa, Wakoshi, Saitama 351-0198, Japan\\
              \email{pg@crab.riken.jp}
         \and
             Institut f\"{u}r Theoretische Physik und Astrophysik, Christian-Albrechts-Universit\"{a}t zu Kiel, Leibnizstr. 15, 24098 Kiel, Germany
         \and
             Zentrum f\"{u}r Astronomie, ITA, Universit\"{a}t Heidelberg, Albert-Ueberle-Str. 2, 69120 Heidelberg, Germany
	 \and
	     Max-Planck-Institut f\"{u}r Radioastronomie, Auf dem H\"{u}gel 69, 53121 Bonn, Germany
         \and
             European Southern Observatory, Alonso de Cordova 3107, Casilla 19001, Santiago, Chile
         \and
             Istituto Nazionale di Astrofisica (INAF) - Osservatorio Astronomico di Bologna, via Ranzani 1, 40127 Bologna, Italy
         \and
             Dipartimento di Astronomia, Universit\`{a} degli Studi di Bologna, Via Ranzani 1, I-40127 Bologna, Italy
             }

   \date{Submitted 2008 Nov 18; Accepted 2009 Feb 13}

  \abstract
   {We present new photometry of 16 local Seyferts including 6 Compton-thick sources in $N$-band filters around 12-\micron, obtained with the VISIR instrument on the 8-m Very Large Telescope. The near-diffraction-limited imaging provides the least-contaminated core fluxes for these sources to date. Augmenting these with our previous observations and with published intrinsic X-ray fluxes, we form a total sample of 42 sources for which we find a strong mid-infrared:X-ray (12.3 \micron:2-10 keV) luminosity correlation. Performing a physically-motivated sub-selection of sources in which the Seyfert torus is likely to be best-resolved 
results in the correlation \lmir $\propto$ \lx$^{1.11\pm 0.07}$, with a reduction of the scatter in luminosities as compared to the full sample. Consideration of systematics suggests a range of 1.02--1.21 for the correlation slope.
The mean 2-keV:12.3-$\mu$m spectral index (\alphaix) is found to be --1.10\p0.01, largely independent of luminosity. Observed 12-$\mu$m bolometric corrections range over $\approx$ 10--30 if a known luminosity trend of intrinsic X-ray bolometric corrections is assumed. Comparison with \iso\ data spanning a larger luminosity range suggests that our correlation can be extended into the quasar regime. The fact that unobscured, obscured and Compton-thick sources all closely follow the same luminosity correlation has important implications for the structures of Seyfert cores. The typical resolution-limit of our imaging corresponds to $\sim 70$ pc at a median $z=0.01$, and we use the tightness of the correlation to place constraints on the dominance of any residual emission sources within these physical scales; we infer an upper-limit of $\approx$40\% of the unresolved flux for any contaminating star-formation, on average. 
We suggest that uncontaminated mid-IR continuum imaging of AGN is an accurate proxy for their intrinsic power. 
}

   \maketitle
%

\section{Introduction}

Typical broad-band spectra of active galactic nuclei (AGN) are known to show a significant bump in the infrared regime (e.g. \citealt{barvainis87}, \citealt{sanders89}, \citealt{elvis94}, and references therein). This is thought to be due to absorption of intrinsic AGN ultraviolet, X-ray and optical radiation by dusty clouds in the AGN \lq torus\rq\ on $\sim$pc-scales, which subsequently re-radiate this energy at infrared (IR) frequencies. But isolating the AGN emission from other ubiquitous contaminating sources, such as bright starbursts within the inner few kpc of AGN host galaxies \citep[e.g. ][]{rodriguezespinosa87,norman88,gonzalezdelgado01}, is difficult due to the finite resolving powers of IR observatories. Telescopes with improving sensitivity, wavelength coverage and resolving power have changed the situation steadily over the past two decades \citep[e.g., ][]{sanders89, mulchaey94, spinoglio95, krabbe01, lutz04, buchanan06, ramosalmeida08_erratum, mushotzky08, melendez08}. These works find increasingly-stronger correlations between the measured IR powers with the intrinsic (X-ray, bolometric or various forbidden emission line) AGN luminosities, but still with significant scatter. 
How much of this scatter is inherently due to the physical conditions of the AGN and torus clouds, as compared to observational selection effects, remains an important unresolved issue.

In two papers \cite[][ hereafter Paper I and Paper II, respectively]{horst06, horst08}, some of us presented high-resolution imaging of nearby AGN with the \vlt/VISIR mid-infrared camera \citep{lagage04} in filters covering the 8--13 \micron\ atmospheric transmission window. The diffraction limit of the 8.2m-diameter mirrors of the \vlt\ over this $N$-band window is $\approx$0\farcs3--0\farcs4, which is well-matched to the mid-IR seeing at the \vlt\ site under reasonable optical seeing conditions; thus, VISIR $N$-band imaging is usually diffraction-limited. A sample of about 30 local AGN were imaged in the two mentioned papers. Due to the unprecedented angular resolution, the unresolved core fluxes are the best (i.e. least biased by extra-nuclear emission) ones available for these objects. 
In a follow-up work, \citet[][ hereafter Paper III]{horst09} compared VISIR multi-filter photometric spectral energy distributions (SEDs) with sensitive \spitzer\ IRS spectra of the same sources and quantified the level of contamination by extended emission, finding non-negligible contamination in \spitzer\ observations of about 40\% of cases.

By extracting intrinsic X-ray luminosities ($L_{\rm 2-10\ keV}$, hereafter \lx) from the recent literature, we were able to compare these with the IR powers measured and found a good correlation between these quantities (see, e.g., Fig. 1 in Paper II). The main results of our works are that {\sl i)} the dispersion in the mid-IR:X-ray correlation is smaller than found in previous works and {\sl ii)} both Seyfert-1s and Seyfert-2s closely follow the same correlation. This second result was not predicted by radiative transfer models in which intrinsic AGN emission is re-processed by obscuring tori with smoothly-distributed dust. This is because an optically-thick line-of-sight through the torus will preferentially show cooler dust and a lower mid-infrared luminosity for the same \lx\ as compared to an optically-thin one (see, e.g., Fig.~1 of \citealt{pierkrolik93}). Result {\sl ii)} above was 
also found in previous studies with lower-resolution ground and space observatories \citep[e.g., ][]{krabbe01, lutz04}, but was ascribed to the expected differences being mitigated due to the presence of unresolved contamination and other selection biases. On the other hand, we found a similar result in our high-resolution \vlt\ imaging, even though contamination is much less severe. This suggested, instead, that dust clouds have significant clumpiness in the AGN tori \citep[e.g., ][]{dullemond05, hoenig06, nenkova08a, nenkova08b, schartmann08} which leads to the observed mid-IR emission from hot illuminated cloud faces in the inner torus regions being largely isotropic. 
The extent to which the mid-IR emission is isotropic can be determined by quantifying the scatter and offsets of obscured and unobscured AGN at any given \lx\ in the mid-IR:X-ray correlation. Our study was so far restricted to sources with low obscuring column densities of cold gas. But the differences between the expectations of clumpy and smooth torus distributions increase with line-of-sight optical depth, so it is important to obtain high-resolution mid-IR imaging of highly obscured sources. 

Even irrespective of the question of torus physical structure, an accurate census of local Seyferts in the mid-IR is important in order to determine their Bolometric accretion energetics and to disentangle AGN from starburst emission. If, after removal of all selection biases, the scatter in the correlation between IR and intrinsic AGN power can be reduced, then the hope is that mid-IR observation may be an independent isotropic indicator of AGN energetics, similar to forbidden \oiii~\l5007\AA\ emission line, far-infrared, hard X-ray or low-frequency radio observations (cf. \citealt{mulchaey94}). 

In this paper, we present new high-resolution mid-IR fluxes around 12-$\mu$m of 16 local Seyferts, including 7 sources obscured by high Compton-thin absorption (with \nh$>$10$^{23}$ cm$^{-2}$) and 6 Compton-thick ones (\nh$>$1.5$\times$10$^{24}$ cm$^{-2}$). Adding literature observations of another well-studied Compton-thick source NGC~1068, as well as our previous sample of 25 detections from Paper II, and comparing these unbiased core mid-IR measurements with the best \lx\ values from the literature, we find the luminosity relationship to be very tight, especially for \lq well-resolved\rq\ targets. Various implications for the core structures of nearby AGN are discussed and useful broad-band (mid-IR to X-ray) spectral relations are derived. A flat Universe with a Hubble constant $H_{\rm 0}=71$ km s$^{-1}$ Mpc$^{-1}$, $\Omega_{\Lambda}=0.73$ and $\Omega_{\rm M}=0.27$ is assumed. 
All literature measurements have been converted to this Cosmology.


\section{Observations}

A sample of 16 targets was observed with VISIR during service nights spanning the semester from 2007 October to 2008 March. These were mainly selected from sources that had been observed (or had planned forthcoming observations) with hard X-ray satellites including \suzaku, \integral\ and \swift, so that their broad-band X-ray spectra could be modeled and 2--10 keV AGN luminosities accurately determined; i.e. the sample is not complete. 
The sky conditions were required to be clear with 0\farcs8 optical seeing or less, during ESO period 80. Observational and data reduction procedures are very similar to those described in Paper II; only minimal details are stated here. 
A standard parallel chopping and nodding technique with a chop throw of 8\arcsec\ was used. Observations were carried out with the small field camera (0\farcs75/pixel) in one or more narrow and intermediate-band filters covering the $N$-band, for each source. As we are interested in determining source continuum luminosities, filters were selected to avoid emission lines and to include continuum regions free of the Silicate absorption trough, depending on source redshift.

The science and standard photometric star calibration frames were reduced with a dedicated pipeline \citep{pantin05_visir}, including removal of detector features and background subtraction. 
The count rate for one full exposure was calculated as the mean of the 3 beams from all nodding cycles of this exposure. In order to minimise the effect of residual sky background, we chose relatively-small apertures ($\approx 10$ pixels = 1\farcs27) for photometry of most sources and corrected the obtained count rates using the radial profiles of standard stars. For a handful of faint targets, profile-fitting (based on the profiles of bright standard-stars) on binned images was carried out instead. As an initial error estimate, the standard deviation of the fluxes amongst the three nodding beams was used, except for a few cases where the statistical error on source counts turned out to be larger. The absolute flux uncertainty is dependent on the counts--to--flux \lq conversion\rq\ factors determined from standard star observations and the known accuracy of stellar templates. For the PAH2 filter, the conversion factor has been found to be constant to within $\approx$ 5\% over a whole observing period, and this minimum uncertainty is added in quadrature to the initial error estimate for all sources. 

The X-ray data used in this paper have been carefully selected (or computed) from recent published studies including spectral modeling over broad energy ranges where available. The \lx\ values are corrected for obscuration, complex spectral components and redshift effects, and should be a measure of the true, intrinsic 2--10 keV AGN power responsible for heating torus clouds. Of course, in cases of extreme obscuration or complexity, these corrections can have large uncertainties -- see \S~\ref{sec:comptonthick}. In this respect, it will be useful to compare against hard X-ray luminosities measured above 10 keV (cf. \citealt{mushotzky08}) in future work, as these should be less susceptible to such complexities. Since data from a variety of published sources is compiled for each source, the error assignment on \lx\ is necessarily somewhat loose. As in Papers I and II, the uncertainty is based on the range of peak-to-peak published values, which should include source variability and other biases. Each source is detailed in the Appendix. 


\section{Results}
\label{sec:results}

All 16 sources were detected in filters around 12-\micron, and the images were dominated in each case by a single IR AGN core. The presentation of images and discussion of full multi-filter $N$-band properties is not the purpose of the present work. We refer any reader interested in the general high spatial-resolution Seyfert morphologies to our Paper III, which includes our entire previous sample. For the purpose of the IR:X-ray luminosity correlation (cf. Papers I and II), the 12.3-\micron\ source flux is the important quantity. We therefore concentrate mainly on the filter most appropriate for probing the continuum closest to 12.3 $\mu$m. Table~\ref{tab:fluxes}\footnote{The full data set will be available from CDS; currently: http://cosmic.riken.jp/pg/gandhi09.dat} lists these filters (referred to as primary filters), the corresponding source fluxes as well as 
monochromatic luminosities at the central filter wavelength ($\lambda L_{\lambda}$, hereafter \lmir). We assume a flat mid-IR SED in these units, i.e. no $k$-correction to 12.3 $\mu$m; instead, 
$\lambda L_{\lambda}^{12.3}$ of a source which has an SED flat in $L_{\lambda}$ units, say, will be underestimated by only 12 per cent (0.05 dex) in the worst case. In addition to our measurements, we included literature fluxes of NGC~1068 \citep{galliano05_ngc1068}, as it is one of the best-studied of nearby AGN. Some secondary filter fluxes are also listed in the table, which we return to in \S~\ref{sec:filter_systematics}.

\begin{figure*}
  \begin{center}
    \includegraphics[width=11.5cm,angle=90]{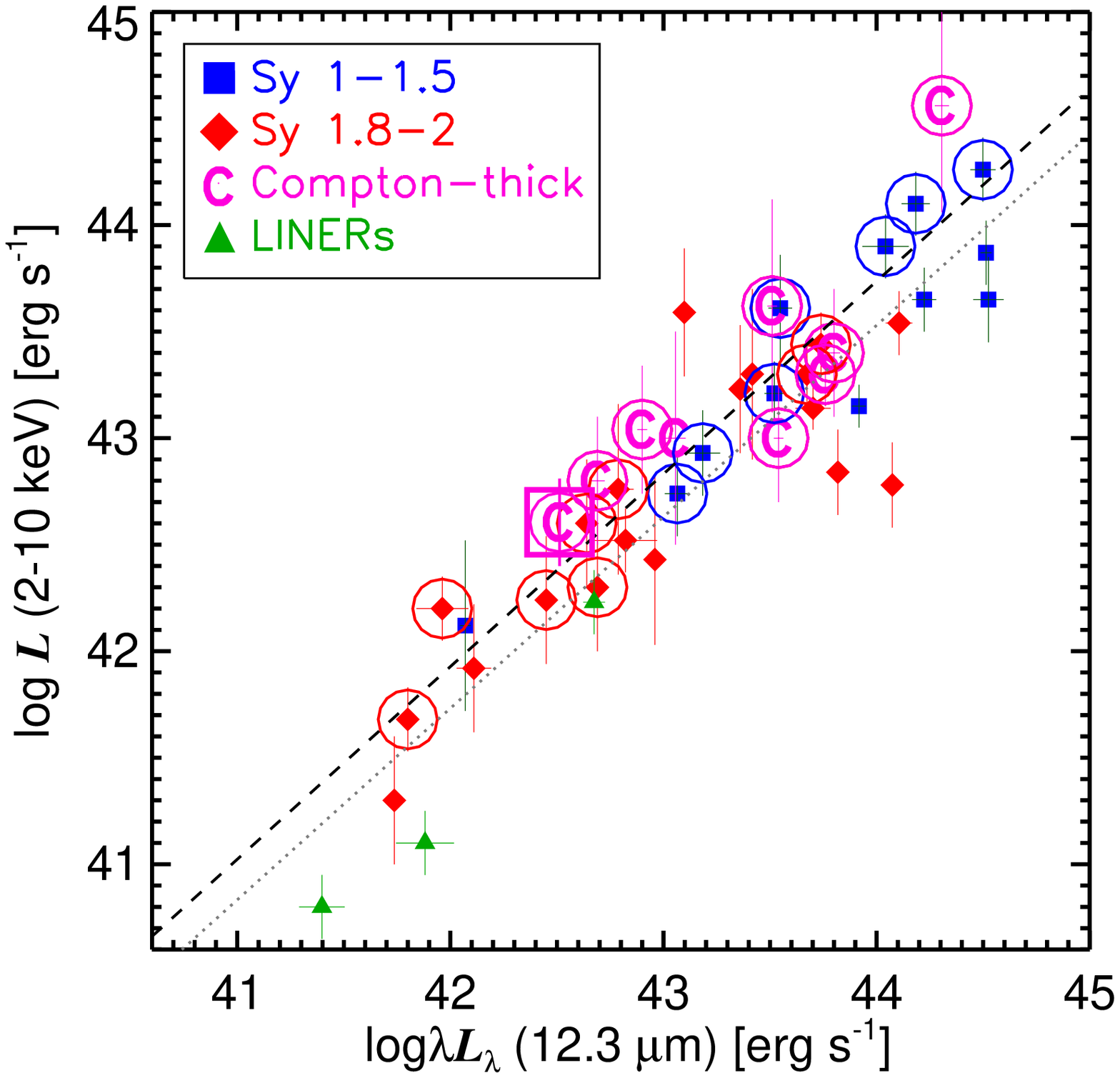}
    \includegraphics[width=11.5cm,angle=90]{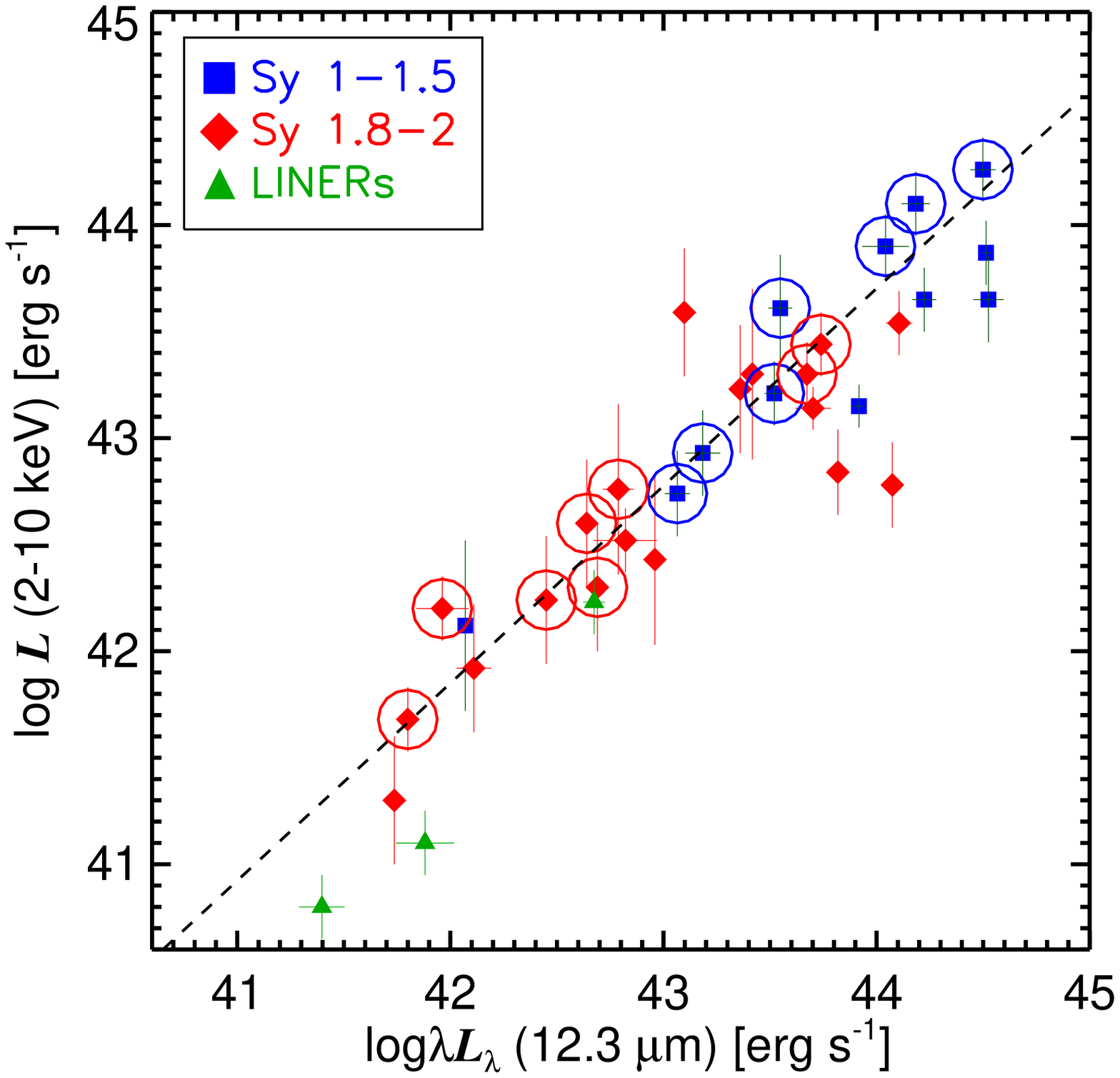}
    \caption{{\em (Top):} Mid-infrared:X-ray luminosity correlation for the sample of 42 VISIR detections from Table~\ref{tab:fluxes} and Paper II. One additional object plotted (to make 43 in all) is ESO~428--G014 (the \lq C\rq\ with both a box and a circle around it), which is not part of our fit sample, but used to validate the correlation in the Compton-thick regime (see \S~\ref{sec:comptonthick}). The sources classified as being \lq well-resolved\rq\ are encircled, and the diagonal dashed line is the fitted correlation for these (Eq.~\ref{eq:correlation} and Table~\ref{tab:correlationfits}). The dotted grey line is the correlation fit to all 42 sources. {\em Bottom):} \lmir--vs.--\lx\ for the 34 Compton-thin sources only, with the correlation fitted to 15 well-resolved sources (Table~\ref{tab:correlationfits}).
    \label{fig:correlation}}
  \end{center}
\end{figure*}

Figure~\ref{fig:correlation} shows the correlation between \lx\ and \lmir. All 25 VISIR detections from Paper II and the 17 measurements from Table~\ref{tab:fluxes} are plotted and split into Seyfert 1--1.5 (12 objects, hereafter \lq Sy 1\rq), Seyfert 1.8--2 or 1h (19 objects, hereafter \lq Sy 2\rq), LINERs (3 objects) and Compton-thick (8 objects) AGN. The \lq 1h\rq\ sub-class of Sy 2s refers to sources with broad optical emission detected in polarized light \citep{veron06}. Since Compton-thick sources (with \nh$\ge$1.5$\times$10$^{24}$ cm$^{-2}$) are of special interest, we chose to plot them with a separate symbol, though all of them fall under the optical \lq Sy 2\rq\ category. In addition to these 42 sources, one additional Compton-thick object (ESO~428--G014) is plotted with a box around it. This is not part of the correlation fit, but is used an an aposteriori check; it will be discussed in \S~\ref{sec:comptonthick}.

As discussed in Paper II, and shown in Table~\ref{tab:fluxes}, the sample covers a wide range (about a factor 40) of spatial scales at the mid-IR angular resolution limit ($\theta_0$) of the \vlt. In order to make a physically-meaningful selection, \cite{horst08} chose only those sources in which the resolved scale ($r_0$), in units of the size of the dust sublimation radius ($r_{\rm sub}$), lay below a certain threshold. By comparing \lmir\ to \lx\ over the range of resolved scales, they identified a value $r_0/r_{\rm sub}=560$ beyond which the ratio \lmir/\lx\ exhibited a sharp increase in its mean value and dispersion. A constant $\theta_0$=0\farcs35 was assumed, and the relation 

\begin{equation}
r_{\rm sub} = 0.5 \sqrt{L_{\rm Bol}^{45}}\ {\rm pc} 
\label{eq:rsub}
\end{equation}

\noindent
was used, where $L_{\rm Bol}^{45}$ is the source Bolometric luminosity in units of 10$^{45}$ erg s$^{-1}$. This relation is approximately correct for a sublimation temperature of 1500 K of graphite grains with an average radial size of 0.05 \micron\ (e.g., \citealt{barvainis87}, \citealt{kishimoto07}; we have also assumed that the bulk of the bolometric contribution is due to ultraviolet photons). All sources with $r_0/r_{\rm sub}$ less than this threshold were classified as being \lq well-resolved.\rq\ In other words, the threshold is a simple multiple of the expected physical size of the torus, and the suggestion is that the rest of the \lq less-resolved\rq\ sources are likely to have non-negligible contamination from extended emission that falls within the central diffraction Airy ring on the VISIR images.

In this paper we use the same source classification criterion for consistency; we will discuss its limitations and the effects of different assumptions in \S~\ref{sec:systematics}. We thus find 22 (of the total 42) sources to be well-resolved, including all Compton-thick sources except NGC~5135. Performing a least-square fit on this sample with a simple power-law (i.e. a linear relation of the form log\lmir$=a+b$log\lx\ in log-space) gives the following best-fit correlation:

\begin{equation}
  {\rm log} \left( \frac{L_{\rm MIR}}{10^{43}} \right)\ =\ (0.19 \pm 0.05)\  +\  (1.11 \pm 0.07)\ {\rm log} \left( \frac{L_{\rm X}}{10^{43}} \right)\\
  \label{eq:correlation}
\end{equation}

\noindent
where the luminosities are normalized to $10^{43}$ erg s$^{-1}$, the approximate mean power of the sample. The regression accounts for luminosity errors on both axes and the uncertainties are based on $\Delta \chi^2=1$ for each best-fit parameter (cf. {\tt fitexy} procedure in \citealt{numericalrecipes}). This correlation curve is over-plotted in Fig.~\ref{fig:correlation}. 
Eq.~\ref{eq:correlation} is completely consistent with (but slightly tighter than) that derived from the smaller sample in Paper II. For easy comparison with that paper, we also state the correlation fit without any normalization, in units of erg s$^{-1}$ directly: 
log\lmir=(--4.37\p3.08)+(1.106\p0.071)log\lx. 
The formal significance of the correlation, according to the well-known Spearman Rank coefficient ($\rho$), is $\rho=0.93$ with a standard Student-$t$ null significance level of $3\times 10^{-10}$. As discussed in Paper II, the correlation is not just a consequence of correlated distances on both axes; the (intrinsic) flux--flux correlation is strong as well, with $\rho=0.86$ at a null significance level of $3\times 10^{-7}$. This can also be seen from a partial correlation analysis between log\lx\ and log\lmir, with the correlation component due to redshift removed; these values are listed in Table~\ref{tab:correlationfits}.

When fitting the correlation to all 42 sources without any selection, the change in parameters is closely equivalent to a simple increase in the intercept on the axis of log\lmir\ by a value of 0.24 (i.e., a mean \lmir\ larger by a factor of 1.7, consistent with inclusion of more IR flux in the cores of the less-resolved sub-sample); see dotted line in Fig.~\ref{fig:correlation}. In this case, $\rho=0.88$ at a null significance level of $1\times 10^{-14}$. On the other hand, excluding the Compton-thick sources has an insignificant effect. A few representative fits to various sub-sample populations are listed in Table~\ref{tab:correlationfits}.

\begin{table*}
  \centering
  \begin{tabular}{lccccr@{~~~}l@{~~~}r@{\p}lr@{~~~}l@{~~~}r@{\p}lcr}
    \hline
    Target                   & Sy  & $z$     &\lognh & $r_0$  & \multicolumn{4}{c}{Secondary filter flux}              & \multicolumn{4}{c}{Primary filter flux}              & log $\lambda L_{\lambda}$  & log \lx            \\
                             &     &         & cm$^{-2}$ & pc  & Name & $\lambda_C$  & \multicolumn{2}{c}{mJy}          & Name & $\lambda_C$        & \multicolumn{2}{c}{mJy} & erg s$^{-1}$               & erg s$^{-1}$ \\
      (1)                    & (2) &  (3)    &  (4)   & (5)     & \multicolumn{2}{c}{(6)}  &  \multicolumn{2}{c}{(7)} &  \multicolumn{2}{c}{(8)}   &  \multicolumn{2}{c}{(9)} &         (10)              &         (11)                \\
    \hline
    {\sl NGC~1068}           & 1h & {\sl 14.4$^d$}&{\sl $>$25} &{\sl 25} & \multicolumn{4}{c}{--}                     & \multicolumn{2}{c}{$\approx$ {\sl 12 $\mu$m}$^{*}$}& \multicolumn{2}{c}{{\sl $\sim 9500$}}& {\sl 43.80\p0.02}        &    {\sl 43.4\p0.3}     \\
    Swift~J0138.6--4001      & 2   & 0.0252  & 23.7   & 175    & SIV  & 10.49              & 20&3                     & PAH2 & 11.25              & 34&3                     &  43.10\p0.04             &        43.6\p0.3        \\
    Swift~J0601.9--8636      & 2   & 0.0062  & 24.0   & 45     & PAH2 & 11.25              & 13&4                     & NEII\_2 & 13.04              & 72&15                     &  42.11\p0.08             &        41.9\p0.3        \\
    ESO~209--G012            & 1.5 & 0.0405  & $<$22  & 280    & PAH2 & 11.25              & 162&10                   & NEII & 12.81              & 188&26                   &  44.22\p0.06             &        43.7\p0.2        \\
    NGC~3081                 & 1h  & 0.0078  & 23.8   & 55     & PAH2 & 11.25              & 161&11                   & NEII\_2 &  13.04             & 138&11                   &  42.64\p0.03             &        42.6\p0.3        \\
    ESO~263--G013            & 2   & 0.0333  & 23.6   & 225    & SIC  & 11.85              & 45&4                     & NEII & 12.81              & 46&3                     &  43.42\p0.02             &        43.3\p0.4        \\
    NGC~3281                 & 2   & 0.0107  & 24.3   & 75     & PAH2 & 11.25              & 481&24                  & NEII\_2 &  13.04             & 1016&52                   &  43.76\p0.02             &        43.3\p0.2        \\
    NGC~3393                 & 2   & 0.0125  & 24.2   & 90     & SIV  & 10.49              & 36&2                     & NEII\_2 & 13.04           & 94&5                     &  42.90\p0.02             &        43.0\p0.3        \\
    NGC~4388                 & 1h  & 16.7$^d$& 23.4   & 30     & PAH2 & 11.25              & 147&11                   & NEII\_2 & 13.04              & 375&28                   &  42.45\p0.03             &        42.2\p0.3        \\
    IC~3639                  & 1h  & 0.0109  & $>$25  & 75     & PAH2 & 11.25              & 326&17                   & NEII\_2 & 13.04              & 542&29                   &  43.51\p0.02             &        43.6\p0.5        \\
    ESO~323--G032            & 1.9 & 0.0160  & --     & 110    & \multicolumn{4}{c}{--}                               & PAH2 & 11.25              & 61&6                     &  42.96\p0.04             &        42.4\p0.4        \\
    NGC~4992                 & 2   & 0.0251  & 23.7   & 175    & PAH1 & 8.59               & 61&6                     & NEII & 12.81              & 71&4                     &  43.36\p0.02             &        43.2\p0.3        \\
    NGC~5728                 & 1.9 & 0.0094  & 24.3   & 65     & \multicolumn{4}{c}{--}                               & NEII\_2 & 13.04           & 123&10                   &  42.60\p0.03             &        42.8\p0.3        \\
    ESO~138--G001            & 2   & 0.0091  & $>$24.3& 65     & PAH2 & 11.25              & 776&40                   & NEII\_2 &  13.04             & 862&45                   &  43.54\p0.02             &        43.0\p0.3        \\
    NGC~6300                 & 2   & 0.0037  & 23.3   & 25     & PAH2 & 11.25              & 278&14                   & NEII\_2 &  13.04             & 727&38                   &  42.69\p0.02             &        42.3\p0.3        \\
    ESO~103--G035            & 2   & 0.0133  & 23.3   & 95     & SIV  & 10.49              & 243&15                   & NEII\_2 & 13.04           & 622&49                   &  43.74\p0.03             &        43.4\p0.2        \\
    NGC~6814                 & 1.5 & 0.0052  & $<$20.5& 35     & PAH2 & 11.25              & 99&6                     & NEII\_2 & 13.04              & 96&6                     &  42.07\p0.02             &        42.1\p0.4        \\
    \hline
  \end{tabular}
  \caption{Observational parameters of our sample of 16 VISIR detections plus archival data on NGC~1068, tabulated in order of increasing Right Ascension. (2) Seyfert type, mostly according to \citet{veron06}, except for the two Swift targets, ESO~323--G032 and NGC 4992; see Appendix. (3) Redshift $z$. Values marked $^d$ are instead distances in Mpc. (4) Obscuring gas column density in units of log [cm$^{-2}$] taken from literature X-ray spectral fits. (5) Approximate resolved scale limit in pc, for an angular resolution of 0\farcs35. (6) Name and central wavelength in $\mu$m of a secondary filter chosen to test systematic variations in the \lmir--\lx\ correlation (details in \S~\ref{sec:filter_systematics}). (7) Secondary filter flux. (8) Chosen primary filter name and central wavelength in $\mu$m. $^{*}$For NGC~1068, the fluxes are from the high angular resolution (0\farcs2--0\farcs4) Keck and Subaru studies \citep{mason06, tomono01} at slightly-differing wavelengths. (9) Observed source core flux in primary filter. 
(10) log of monochromatic primary filter luminosity expressed in erg s$^{-1}$. For our mid-IR:X-ray correlation we assume a flat SED in $\lambda L_{\lambda}$, i.e. no $k$-correction to 12.3 $\mu$m. (11) log of 2--10 keV intrinsic X-ray luminosity. See Appendix for details on the literature used and notes on individual sources.
    \label{tab:fluxes}}
\end{table*}

\begin{table*}
  \centering
  \begin{tabular}{lcccccr}
    \hline
    Sub-sample                 & $N$       & $\rho_{\cdot z}$              &    $a$      &     $b$    &  $\bar{r}$  &  $\sigma_r$\\
       (1)                     & (2) &             (3)                     &    (4)      &     (5)    &      (6)    &    (7)    \\
    \hline
    {\bf All}                  &  {\bf 42} & {\bf 0.81}                    &  {\bf 0.41\p0.03} & {\bf 1.11\p0.04} & {\bf 0.30} & {\bf 0.36}\\
    All, excluding CT          &  34       &      0.79                     &  0.43\p0.04 & 1.11\p0.04 &     0.35    &    0.36\\
    All Sy~1                   &  12       &      0.71                     &  0.34\p0.09 & 1.18\p0.14 &     0.34    &    0.31\\
    All Sy~2 + CT              &  27       &      0.77                     &  0.40\p0.06 & 1.40\p0.11 &     0.25    &    0.38\\
      & & & & \\  
    {\bf Well-resolved}        &  {\bf 22} & {\bf 0.89}                    &  {\bf 0.19\p0.05} & {\bf 1.11\p0.07} & {\bf 0.15} & {\bf 0.23}\\
    Well-resolved, excluding CT&  15       &      0.93                     &  0.18\p0.05 & 1.08\p0.07 &     0.17    &    0.18\\
    Less-resolved              &  20       &     0.76                      &  0.62\p0.05 & 1.11\p0.06 &     0.47    &    0.40\\
      & & & & \\  
      \multicolumn{7}{c}{\underline{Systematic variations}}\\  
    Well-resolved (Systematic 1$\uparrow$)  & 25 & 0.82                    & 0.25\p0.05 & 1.21\p0.08 &     0.17    &    0.30\\
    Well-resolved (Systematic 1$\downarrow$)& 16 & 0.90                    & 0.26\p0.05 & 1.07\p0.08 &     0.21    &    0.23\\
    Well-resolved (Systematic 2)            & 22 & 0.83                    & 0.22\p0.05 & 1.09\p0.07 &     0.16    &    0.27\\
    Well-resolved (Systematic 3)$^{\dag}$   & *  & *                       & 1.87\p0.31 & 1.11\p0.10 &       *     &      * \\
    Well-resolved (Systematic 4)            & *  & *                       & 0.15\p0.05 & 1.02\p0.07 &       *     &      * \\
    All (Systematic 5)                      & 42 & 0.74                    & 0.42\p0.03 & 1.11\p0.04 &     0.30    &    0.44\\
    Well-resolved (Systematic 5)            & 22 & 0.78                    & 0.21\p0.05 & 1.09\p0.07 &     0.13    &    0.31\\
    \hline
  \end{tabular}
  \caption{Correlation properties between log\lx\ and log\lmir\ to various sub-sample populations (Column 1), each containing $N$ sources (Col. 2). The partial correlation coefficient ($\rho_{\cdot z}$), with the effects of redshift removed are listed in Col. 3. Regression of the form log$L_{\rm MIR}^{43}$= $a$+$b$log$L_{\rm X}^{43}$ (luminosities in units of 10$^{43}$ erg s$^{-1}$) was carried out; the best-fit parameters are listed in Cols. 4 and 5 along with their 1-$\sigma$ uncertainties. Cols. 6 and 7 list the mean value and scatter in $r$=log(\lmir/\lx); see \S~\ref{sec:dispersion}. \lq CT\rq\ refers to Compton-thick sources. The two solutions likely to be of most interest (see \S~\ref{sec:summary}) are written in bold. The final seven rows refer to tests for systematic variations in the correlation; see \S~\ref{sec:scale_systematics}--~\ref{sec:filter_systematics}. The test fit marked with a $^\dag$ (Systematic 3) is carried out in linear (not log) luminosity space (see \S~\ref{sec:errors}). For Systematics 3 and 4, the sub-samples are identical to the main well-resolved sample of 22 sources, and all \lq *\rq\ parameters are identical as well (see bold text under \lq Well-resolved\rq).
    \label{tab:correlationfits}}
\end{table*}


\section{The intrinsic mid-IR emission of AGN cores}

\subsection{Dispersion in \lmir\ vs. \lx}
\label{sec:dispersion}

\begin{figure*}
  \begin{center}
    \includegraphics[width=12.5cm,angle=0]{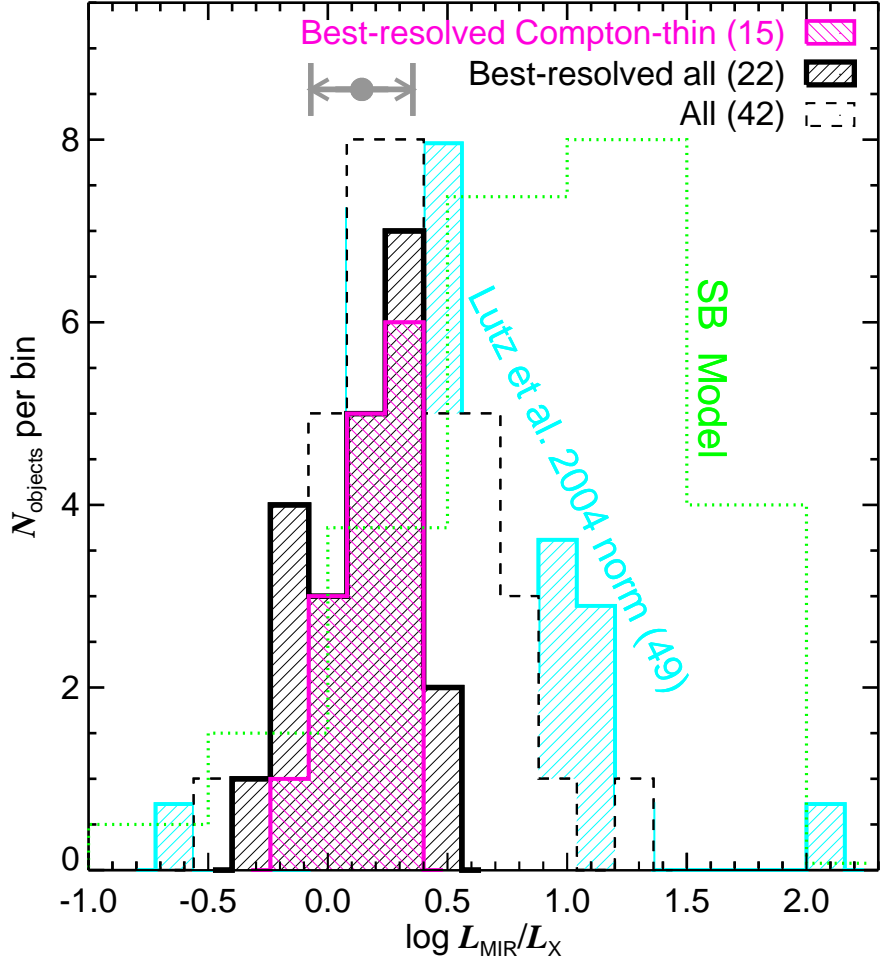}
    \caption{Dispersion in log\lmir/\lx. For our data, three distributions are plotted in order of increasing depth into the figure: \emph{i)} that for 15 well-resolved Compton-thin sources is shown by the pink cross-hatched histogram; \emph{ii)} that for all 22 well-resolved sources by the black hatched area; and \emph{iii)} that for all 42 sources by the unfilled histogram with dashed outline. The grey filled circle and arrow shows the mean and extreme values of the correlation curve in Eq.~\ref{eq:correlation} over the fitted luminosity range. For the sample of \citet{lutz04}, the light-blue hatched area shows the distribution of 49 sources in log ($\lambda L_{\lambda}^{6 \mu{\rm m}}$/\lx), normalized to the maximum value of our full data histogram. Finally, the light-green, dotted histogram is the model prediction for the AGN+starburst model of \citet{ballantyne08}, also normalized.
    \label{fig:hists}}
  \end{center}
\end{figure*}

\lmir\ correlates tightly with \lx\ over three orders of magnitude in Seyfert luminosity. The small scatter is emphasized by plotting the distribution of the log-space luminosity ratio $r$=log(\lmir/\lx) in Fig.~\ref{fig:hists}. Our full dataset of 42 sources has a mean and standard-deviation of ($\bar{r}$, $\sigma_r$)=(0.30, 0.36) in this ratio, while the corresponding values for the 22 well-resolved sources of all types are (0.15, 0.23). 
The 15 Compton-thin well-resolved sources have the least scatter of all with corresponding values of (0.17, 0.18); see Table~\ref{tab:correlationfits} and right-hand plot in Fig.~\ref{fig:correlation}. The dispersion in log\lmir\ with respect to the best-fit correlation curves (i.e. observed vs. predicted log\lmir) in the individual sub-samples are very similar to the above $\sigma_r$ values.

The light blue histogram in the same figure refers to the \iso\ study of \citet{lutz04}, for which we present their distribution in log($\lambda L_{\lambda}^{6 \mu{\rm m}}$/\lx) of 49 AGN -- the ones with 6-\micron\ continuum detections for which intrinsic X-ray flux estimates could be determined. We also removed any sources likely to have significant IR starburst contamination (as defined by these authors). For this distribution, ($\bar{r}$, $\sigma_r$) = (0.53, 0.45). The spread is comparable to (only slightly larger than) the scatter for our full sample of 42 sources. The mean value, on the other hand, is significantly larger than for our sample, which may be either due to residual contamination within the larger \iso\ apertures or to the heterogeneous nature of their sample; we return to this in \S~\ref{sec:quasars}. We can at least confidently rule out this difference being due to their bluer 6 \micron\ central wavelengths. Using median mid-IR AGN SEDs presented by \citet{hao07}, we find that $\lambda L_{\lambda}^{6 \mu{\rm m}}$ for Sy 1 (Sy 2) is about 19 (54) per cent {\em lower} than $\lambda L_{\lambda}^{12.3 \mu{\rm m}}$, which is all due to red continuum shapes and not due to contamination by PAH emission or Silicate absorption. Thus wavelength differences alone should have resulted in a lower mean log($\lambda L_{\lambda}^{6 \mu{\rm m}}$/\lx) than that for our sample. If we do convert the \citeauthor{lutz04} luminosities to rest-frame $\lambda L_{\lambda}^{12.3 \mu{\rm m}}$ using the Sy 1, Sy 2 and quasar $k$-corrections from \citeauthor{hao07}, we find ($\bar{r}$, $\sigma_r$) = (0.71, 0.50).

With regard to the smaller dispersion in our well-resolved sub-sample, is this simply due a serendipitous selection of a few sources from our full sample which, in reality, have no physical inter-relation? The dispersion of a smaller sample may well be measured to be smaller due to statistical, rather than physical reasons, as pointed out by \citet{lutz04}. We rule out this possibility based on two arguments. First, note the clear and large differences between the scatter of the well-resolved sub-sample and of other sub-samples containing approximately equal objects in Table~\ref{tab:correlationfits}. The difference is especially clear when comparing against the less-resolved sources which have much larger $(\bar{r}$, $\sigma_r$)=(0.5, 0.4) values. Next, in order to further estimate the effect of any fortuitous selection, we constructed 30000 bootstrap populations of 22 sources each, drawn randomly from the full sample of 42 sources without replacement. We measured the dispersion in each of these populations and computed the fraction of random populations with a dispersion less than or equal to the dispersion measured in the data. The result is that our measured data dispersion is robust to a serendipitous selection at 99.2 per cent; only about 250 random sub-samples had a variance smaller than that of the data. 

\subsection{The compositions of the unresolved mid-IR cores}
\label{sec:corecomposition}

The imaged VISIR cores of the vast majority of our full sample appear completely unresolved, even at the high-resolution of the VLT (as mentioned above; see also extensive discussion in Paper III). The physical resolution limits ($r_0$) of our observations range over tens to hundreds of pc (see Table~\ref{tab:fluxes}; Paper II). This is larger by factors of several hundred than an inner torus radius computed according to Eq.~\ref{eq:rsub}. The full radial extent of the torus, on the other hand, is more difficult to constrain (e.g. \citealt{granato97}). But our resolution limit is still larger by factors of a few to several tens than the outer radii suggested by recent clumpy torus models (cf. \citealt{hoenig06}, \citealt{nenkova08b} and references therein). So our listed $r_0$ values should be taken as strong upper limits of the physical sizes of the individual tori (at least the size relevant for the distribution of hot mid-IR--emitting dust).

What about the composition of the unresolved AGN cores? The small scatter in our correlation can be used to constrain their average composition to some extent because any contribution by multiple, unresolved components is bound to increase the dispersion in log(\lir/\lx). 

A very tight correlation may argue for a single population of particles emitting both X-ray and IR photons via some non-thermal process. Though we cannot discount an important non-thermal mid-IR component for every one of our sources, such a component is not the dominant cause of the tight correlation, as can be argued in several ways. Only a few of our targets show significant non-thermal emission in the radio, including 3C~445 and Cen~A (in which a non-thermal mid-IR synchrotron component has indeed been inferred to exist by \citealt{meisenheimer07}; though see \citealt{radomski08} for other possibilities). 
Furthermore, the X-ray power-laws, if extended to the IR regime, severely underestimate the observed VISIR fluxes in most cases (but see \citealt{carleton87}). Finally, extensive modeling of individual sources also supports the weakness of any non-thermal component. An extended mid-IR structure has been resolved out with interferometric observations \citep{jaffe04} in at least one of the objects, NGC~1068. \citet{hoenig08_ngc1068_nonthermal} have modeled the broad-band emission from this source and placed strict limits on the non-thermal radiation, favouring instead thermal emission from a clumpy torus cloud medium. 

In any case, our empirical luminosity correlation is valid and important, regardless of its underlying physical origin. We leave further discussion of this issue to future works including detailed monitoring observations and modeling of AGN tori; instead, we focus now on another consequence of the small correlation spread -- limits on the spatial extent and power of star-formation (SF) in the mid-IR at high spatial resolution.

\subsection{Constraining unresolved mid-IR star-formation}
\label{sec:starbursts}

Powerful nuclear starbursts (SB) should produce copious amounts of infrared emission, but comparatively little X-rays \citep[e.g. ][]{rodriguezespinosa87,ranalli03}. While the bulk of this is radiated at far-IR wavelengths (at $\ga$60-\micron), the mid-IR flux around 10-\micron\ due to SF is far from negligible (e.g. \citealt{maiolino95_midIR}, \citealt{galliano05_starbursts}, \citealt{siebenmorgen08}, our Papers II and III). Furthermore, the instantaneous IR luminosity of a SB can vary by more than an order of magnitude \citep[e.g. ][]{cidfernandes04, davies07}, depending on its size and age. The resultant expected scatter in the combined AGN+SB power can be large. On the other hand, our tight mid-IR:X-ray correlation suggests that AGN emission in the point cores of our high-resolution imaging completely dominates over any SF on similar scales at mid-IR wavelengths, and provides us the opportunity to place limits on any unresolved SB contribution.

In a recent work, \citet{ballantyne08} investigated the efficacy of nuclear SBs at $z\sim 1$ to obscure the sources that power the bulk of the cosmic X-ray background radiation. Although the physical motive and cosmic epoch of Ballantyne's study is different from ours, his numerical models make useful predictions of the emitted mid-IR (from the SB) and X-ray (AGN) fluxes, which can be compared with our observations. In particular, the expected distribution of log($\lambda L_{\lambda}^{6 \mu{\rm m}}$/\lx) from his Fig.~11 is illustrated with the green histogram in Fig.~\ref{fig:hists}. For this modeled distribution, $\sigma_r$$\approx$0.75, a factor of about 3 (2) greater than our sample of well-resolved (all) sources, though it must be noted that the nuclear SB luminosity ratios are indicative values only, according to \citeauthor{ballantyne08}. Conversion of the model luminosities to 12 \micron\ should not affect the dispersion much (cf. Fig.~7 in \citealt{ballantyne08}, in which the 6 and 12 \micron\ SB luminosities are approximately equal).

\begin{figure}
  \begin{center}
    \frame{\includegraphics[width=8cm,angle=0]{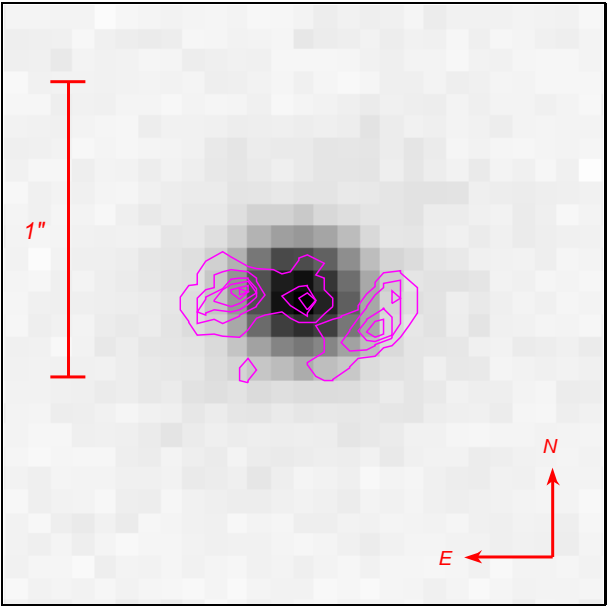}}
    \caption{VISIR NEII\_2 ($\lambda_C=13.04$-\micron) image of IC~3639, showing the clean point-like core profile. North is up and East is to the left; the image was obtained at $0\farcs 75$/pixel sampling and is $2\arcsec \times 2\arcsec$ wide. 
The overlaid contours are from the ultraviolet (UV) Faint Object Camera image obtained from the \hst\ archive. Contour levels range from 10\% to 100\% of the peak UV flux. The extended SB emission over the central arcsec, especially the central \lq horizontal-S\rq\ shaped structure of knots \citep[cf. ][]{gonzalezdelgado98} are clearly visible in the UV, but not at 12-\micron.
    \label{fig:ic3639}}
  \end{center}
\end{figure}

SF is certainly present around the cores of Seyferts, but it is the spatial scale being probed that is the important quantity here. As an illustrative example of this, in Fig.~\ref{fig:ic3639} we show our VISIR high-resolution imaging of one target -- IC~3639 -- whose nuclear SB was studied in detail by \citet{gonzalezdelgado98} and found to have a bolometric luminosity (determined mainly from ultraviolet observations) very similar to the total power emitted by the central AGN. The point-like mid-IR profile is in sharp contrast to the over-plotted contours of \hst\ imaging obtained by these authors. Images of all our targets from Papers I and II can be found in Paper III \citep{horst09}; other multi-band images and SEDs from the present sample will be presented in a future work. 

The median resolved scale ($r_0$) for our 22 targets classified as being well-resolved is 70\p11 pc, and for the rest of the 20 less-resolved targets is 115\p24 pc (errors correspond to the standard deviation of the median of 10000 bootstrap samples drawn with replacement). Another way to gauge the difference between the two samples is a two-sided Kolmogorov-Smirnov (KS) test, according to which the distributions of resolved scales differ at the 97 per cent level.  On the other hand, the distributions of $r_{\rm sub}$ values of the two samples are consistent with each other. In other words, the dominant factor governing source classification turns out to be simply the distance, with the well-resolved sources being on average closer to us (their median redshift is 0.010, while $z_{\rm median}$ for the rest is 0.017) and with smaller $r_0$ values. 

It then follows that the relevant scale within which the AGN completely dominates the SB is the central $\sim 60-80$ pc (the median of the well-resolved sample) of the nuclei of local Seyferts. 
In Table~\ref{tab:correlationfits}, we found $\sigma_r=0.23$ for the well-resolved sample. If this scatter were to be solely due to nuclear SB and assuming negligible X-rays from them, it would imply that SB within these physical scales typically have a 12 \micron\ luminosity of no more than 1--10$^{-\sigma_r}$ = 40\% of the total unresolved (AGN+SB) $L_{\rm MIR}$. Assuming the dispersion for the full sample of 42 sources increases this limit to 56\% while the use of the Compton-thin sub-sample reduces it to 34\%. 
These are conservative upper-limits, because other effects such as AGN infrared variability and the scatter due to statistically-differing distributions of torus clouds will introduce additional uncertainty; so the true constraint will undoubtedly be even tighter.

Beyond a radius of $\sim 100$ pc (using the less-resolved sample alone), the limit on the SB power increases to $\sim 1.5$ times the 12 \micron\ luminosity of the AGN itself, allowing for the bulk of star-formation to occur on these scales. This is completely consistent with other recent studies: e.g. \citet{imanishi02}, \citet{soifer03} and \citet{davies07}. The last authors obtained detailed integral-field near-IR spectroscopy of several local Seyferts and found AGN and nuclear SB powers to be comparable only on kpc scales. 
$K$-corrections of their results to 12-\micron\ are expected to be small compared to the differences due to integration over different spatial scales. Finally, we emphasize that although our observed scatter is much smaller than that presented by \citet{ballantyne08} this is not, by any means, a negation of their model which is relevant at $z\sim 1$. It simply means that we are resolving out any similar SBs at the present epoch, if they exist. 

In summary, we suggest that our high-resolution observations are predominantly probing the nuclear torus emission, unbiased by any overwhelming contamination due to powerful SF. 

\subsection{Infrared:X-ray spectral indices}

Having determined the true mid-IR and X-ray luminosities attributable to the AGN themselves, we can now derive their intrinsic SED shapes. A spectral index connecting two widely-separated frequencies is often used to describe sparsely sampled broad-band spectra. In accordance with the widely-used optical:X-ray spectral index (\alphaox), we define the index 

\begin{equation}
  \alpha_{\rm IX}=\frac{{\rm log} ( L_{2\ {\rm keV}}^{\nu}/L_{12.3\ \mu{\rm m}}^{\nu})}{{\rm log} (\nu_{2\ {\rm keV}}/\nu_{12.3\ \mu{\rm m}})}=0.233\ {\rm log} ( L_{2\ {\rm keV}}^{\nu}/L_{12.3\ \mu{\rm m}}^{\nu})
\label{eq:alphaix}
\end{equation}

\noindent
where $L^{\nu}$ represent the monochromatic luminosities [erg s$^{-1}$ Hz$^{-1}$] at frequencies ($\nu$) corresponding to 2 keV and 12.3 \micron, both in the rest-frame. We have computed $L^{\nu}_{\rm 2\ keV}$ from the individual X-ray photon-indices (usually determined from spectral fits over the $\sim 0.5-12$ keV range) in the literature. 

\begin{figure*}
  \begin{center}
    \includegraphics[width=8.5cm,angle=90]{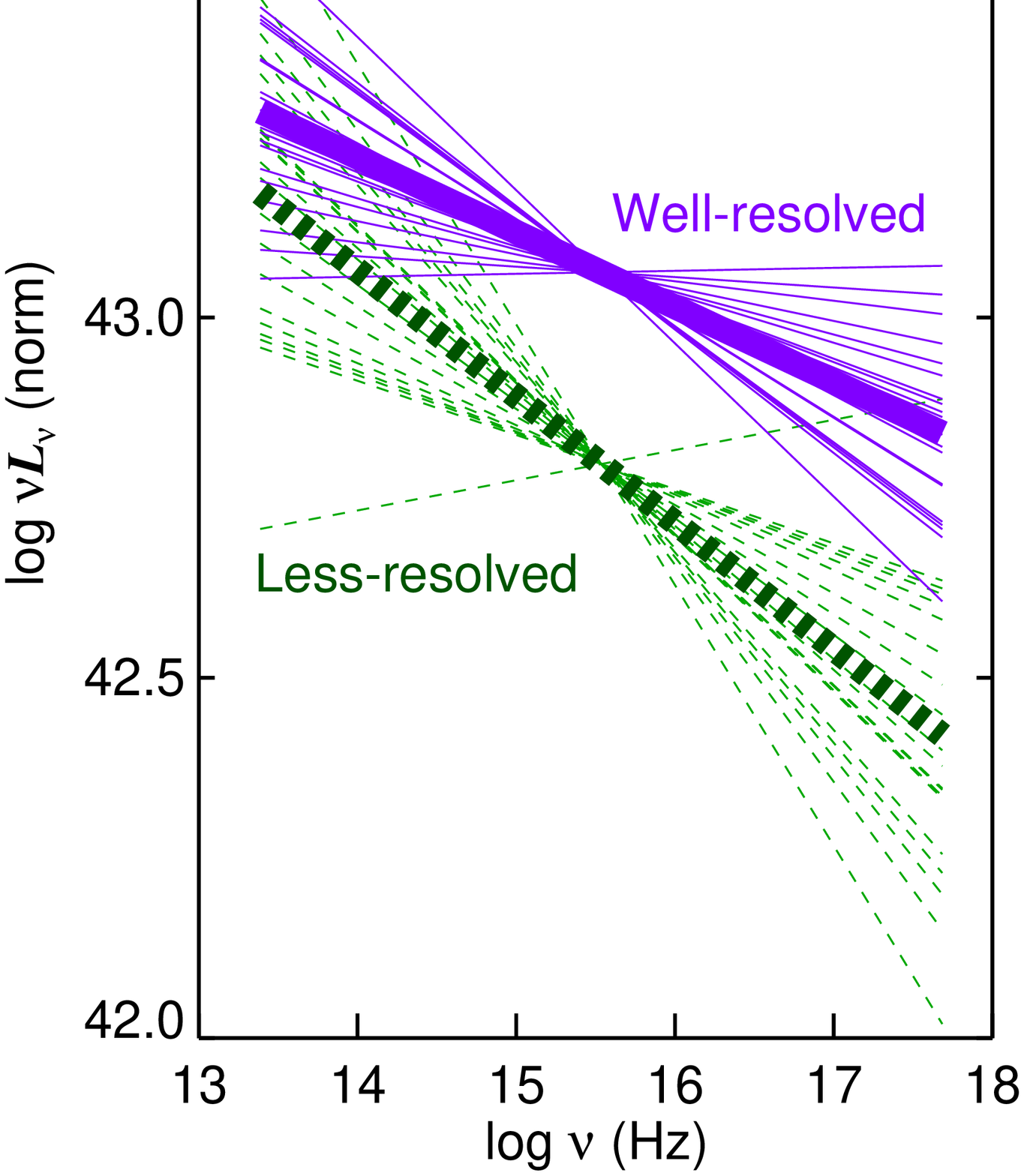}
    \hspace*{-0.5cm}
    \includegraphics[width=8.5cm,angle=90]{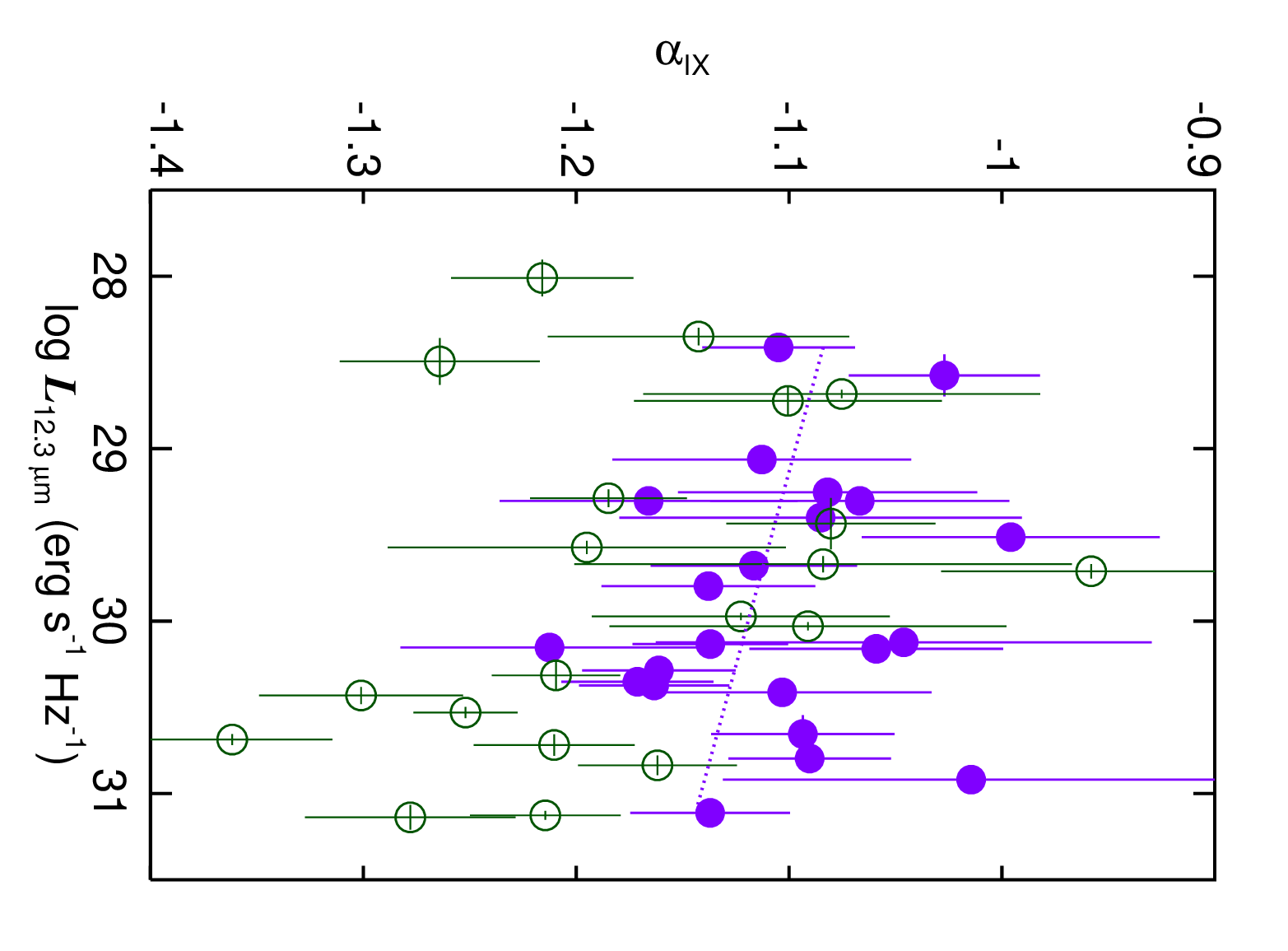}
    \caption{\emph{(Left):} The two-point broad-band SEDs of the well-resolved (purple, solid lines) and less-resolved (green, dashed lines) sub-samples, showing the distribution of \alphaix\ values. Within both sub-samples, sources have been normalized for clarity to the average sub-sample luminosity at the mean frequency of log$\nu$=15.54. The thick central lines denote the mean \alphaix\ slopes. \emph{(Right):} The distribution of individual \alphaix\ values as a function of log $L_{12.3 \mu{\rm m}}^{\nu}$. The dotted line is the best-fit regression curve for the well-resolved sub-sample and has a slope of --0.02.
    \label{fig:alphaix}}
  \end{center}
\end{figure*}

We then find a mean \alphaix=--1.10\p0.01 (0.06) and --1.18\p0.02 (0.10) for the well-resolved, and less-resolved, samples respectively, and --1.14\p0.01 (0.08) for the whole sample; the numbers in brackets list the full scatter ($\sigma_{\alpha_{\rm IX}}$). This distribution is shown in Fig.~\ref{fig:alphaix}. The plot shows that not only is the spectral slope for the less-resolved sources steeper (consistent with including more mid-IR emission for the same \lx), but that the dispersion of slopes for these sources is larger. 

Often in deep surveys, it is not possible to determine the X-ray photon-index from individual spectral fits, due to the faintness of sources. For such cases, we note it is a good approximation to assuming a constant $\Gamma\approx 1.9$ over the 2--10 keV range. This is because the typical observed range of $\Gamma$ is $\approx$ 1.5--2.3 (e.g. \citealt{nandra94, mateos05_lockman, piconcelli05}), and the difference between a constant $\Gamma=1.9$ and the extremes of the stated range results in only a small difference of $\Delta$\alphaix=\p0.04 (assuming no change in \lx). In terms of predicting the luminosity of a source at one frequency from a flux (i.e. luminosity) measurement at the other frequency, using the average spectral slope computed above will give an answer accurate to within a factor of 1.5, over the mentioned range of X-ray photon-indices.

It is well-known that the ultraviolet:X-ray spectral index (\alphaox) of AGN is strongly dependent upon luminosity \citep[e.g. ][]{vignali03_alphaox, steffen06}. The correlation of Eq.~\ref{eq:correlation} is only slightly non-linear, which implies that the corresponding mid-IR:X-ray index should instead depend only weakly on luminosity. Assuming again that a constant X-ray photon-index ($\Gamma=1.9$) is applicable to all sources for simplicity, the expected power-law index of dependence of \alphaix\ on the monochromatic IR luminosity is easily derived from Eq.~\ref{eq:correlation} and Eq.~\ref{eq:alphaix} to be $0.233 (b^{-1}-1)$, where $b$ is the best-fit correlation slope. We found $b=+1.11$ for the well-resolved sub-sample, which implies an expected dependence of \alphaix$\ \propto (L_{\rm MIR}^{\nu})^{-0.02}$. The right-hand panel of Fig.~\ref{fig:alphaix} shows the full dataset of individual sources in the \alphaix-vs-$L_{\rm MIR}^{\nu}$ plane; the errors on \alphaix\ incorporate the uncertainties on both X-ray and mid-IR luminosities. The dotted line shows the regression of the well-resolved sources, and its best-fit slope of --0.022\p0.013 matches the expected value.

\subsection{Bolometric corrections}

Correction factors (\fbol) for obtaining full bolometric AGN power emissions based on observations over limited energy ranges are crucial for computing the total accretion energy release in the Cosmos, as well as studying black hole growth and evolution \citep[e.g. ][]{elvis94, fi, marconi04, hopkins07_bolometricqlf, gilli07}. In particular, it is natural to connect source distributions seen at mid-IR and X-ray energies \citep[e.g. ][]{g03} as these are also good probes of any evolution in obscured AGN populations \citep[e.g. ][]{ballantyne07, treister08}.

Intrinsic bolometric corrections connecting the absorption-corrected X-ray power with the integrated (optical/ultraviolet to hard X-ray) emission have been found to increase with AGN luminosity (e.g. \citealt{marconi04}; see also above discussion on \alphaox). 
If the mid-infrared flux, on the other hand, is the result of reprocessing and thermalization of intrinsic AGN photons within torus dust clouds, then the infrared correction factor will depend upon radiative transfer within each cloud, as well as the clouds covering factor. 

From an empirical standpoint, the expected infrared bolometric corrections (\fbolmir=\lbol/\lmir) can be easily derived by combining our \lmir\ : \lx\ correlation with the $L_{\rm Bol}:L_{2-10}$ trend found by \citet{marconi04}. These last authors present the X-ray bolometric corrections (\fbolx=$L_{\rm Bol}/L_{2-10}$) as a function of the scaled bolometric luminosity $\mathcal{L}$=log(\lbol/\lsun)--12, and we adopt the same units to ease comparison. The value of \lsun\ used is 3.826$\times$10$^{33}$ erg s$^{-1}$. The following equation is obtained: 

\begin{equation}
{\rm log}{\rm f}_{\rm Bol}^{\rm MIR}({\cal L}) = b {\rm log f}_{\rm Bol}^{\rm X}({\cal L}) - a + (1-b)({\cal L}+12+{\rm log}L_{\sun} -43)
\label{eq:lbol}
\end{equation}

\noindent
Expressing it in this form makes clear the dependence on our fitted slope and intercept (which can be picked from Table~\ref{tab:correlationfits} for any of the fits), and the additional log-luminosity units collected at the end. 
It also aids in Monte Carlo determination of confidence intervals, because \citet{marconi04} state an uncertainty of 0.1 in log\fbolx\ and we have computed the uncertainties in $a$ and $b$ in Table~\ref{tab:correlationfits}. Assuming these to be normally-distributed, the resultant 1-$\sigma$ uncertainty can be derived from the scatter in a large number (we used 10000) of randomizations of Eq.~\ref{eq:lbol} at various values of ${\cal L}$. Substituting the best-fit $a$ and $b$ for our well-resolved sample (Eq.~\ref{eq:correlation}), and expanding log\fbolx\ from the relevant Eq.~21 in \citet{marconi04} then gives

\begin{equation}
{\rm log}{\rm f}_{\rm Bol}^{\rm MIR}({\cal L}) = 1.24 + 0.16{\cal L} + 0.012{\cal L}^2 - 0.0015{\cal L}^3
\label{eq:lbol_good}
\end{equation}

\begin{figure*}
  \begin{center}
    \includegraphics[width=12.5cm,angle=0]{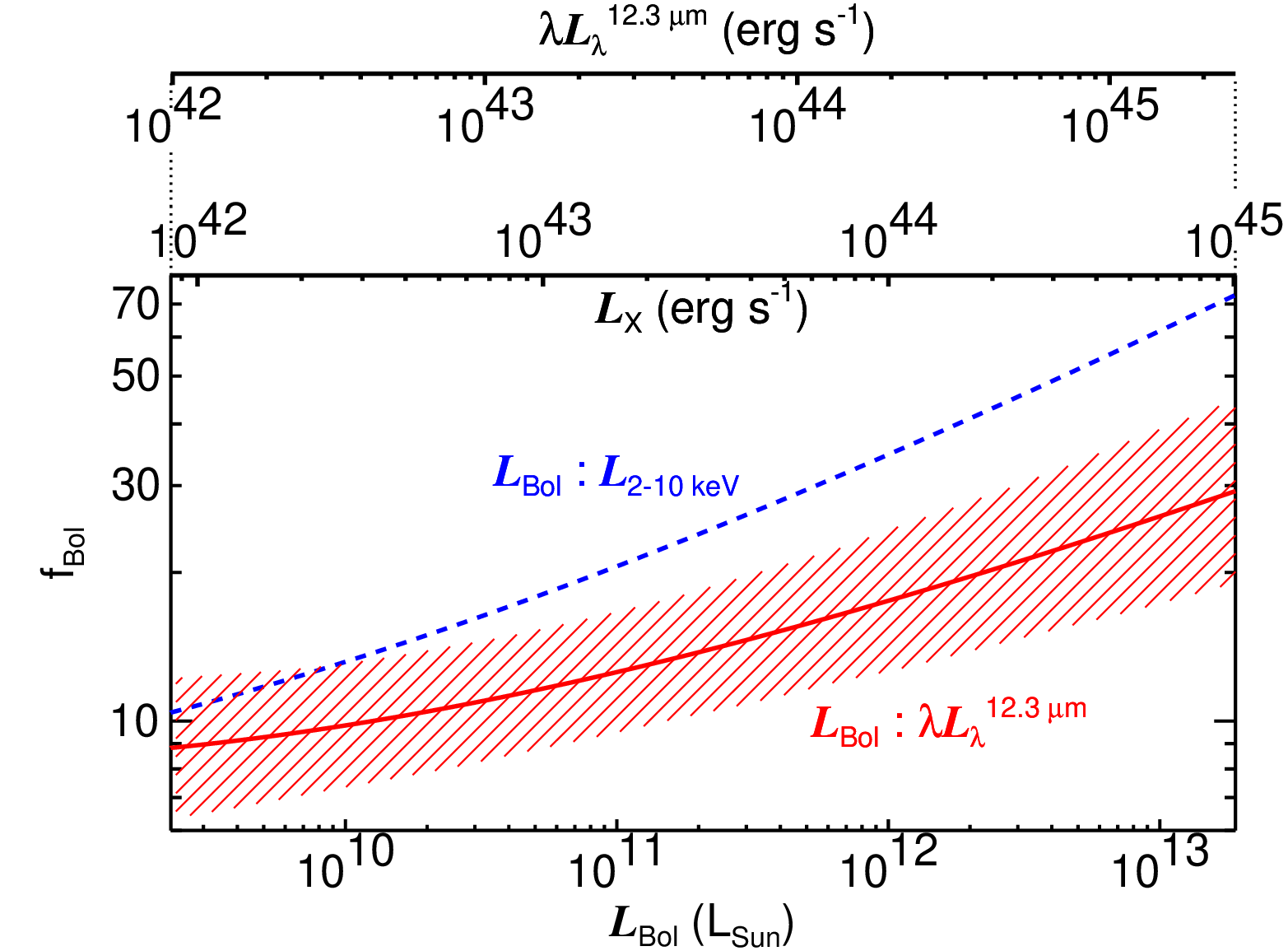}
    \caption{Mid-infrared bolometric correction \fbolmir=\lbol/\lmir\ (red continuous curve) -- determined from our \lmir:\lx\ correlation for well-resolved sources (Eq.~\ref{eq:correlation}) and the \fbolx=\lbol/\lx\ trend observed by \citet{marconi04}. This latter X-ray trend is shown as the blue dashed curve. 
The red hatched area shows the 16th and 84th percentile confidence region in 10000 Monte Carlo randomizations, incorporating normal uncertainties on our correlation parameters as well as on log\fbolx.
    \label{fig:bolcorr}}
  \end{center}
\end{figure*}

\noindent
This curve is drawn in Fig.~\ref{fig:bolcorr}, along with the computed uncertainty. The plot shows a monotonically-increasing f$_{\rm Bol}^{\rm MIR}$ correction for the Sy luminosities that we probe, with a range of average values of $\approx 10-30$, while the X-ray correction is steeper. These values lies at (and above) the upper range of Bolometric corrections found by \citet{spinoglio95}; this is consistent with us probing closer in to the individual nuclei than the \iras\ data used by these authors, but specific differences (e.g. in computation of \lbol, as well as their much larger sample size) need to be kept in mind. 

It is important to note a caveat implied by more recent studies. \citet{vasudevanfabian07, vasudevanfabian08} find only a marginal trend of \fbol\ with \lx, inferring instead a stronger dependence on the the Eddington ratio which is proportional to \lbol\ normalized to the black hole mass. In this case, infrared bolometric corrections may be expected to vary more weakly with luminosity than in Fig.~\ref{fig:bolcorr}, or to show more complex behaviour.

\subsection{From Seyferts to quasars?}
\label{sec:quasars}

Our study has probed local AGN, which are predominantly Seyferts. AGN with quasar-like X-ray luminosities (\lx$>>10^{44}$ erg s$^{-1}$) are absent from our sample due to this selection of nearby targets and because distant sources are less-resolved. On the other hand, it becomes increasing difficult for SF to dominate the energetics of sources at the high end of bolometric luminosities. The bulk of the observed IR fluxes of the most luminous Seyferts and quasars are likely to be due to accretion onto the central black hole, irrespective of instrumental angular-resolution. Thus, we should be able to check our correlation in the luminous quasar regime by using space-telescopes observations of distant quasars.

\begin{figure}
  \begin{center}
    \includegraphics[width=8.7cm,angle=90]{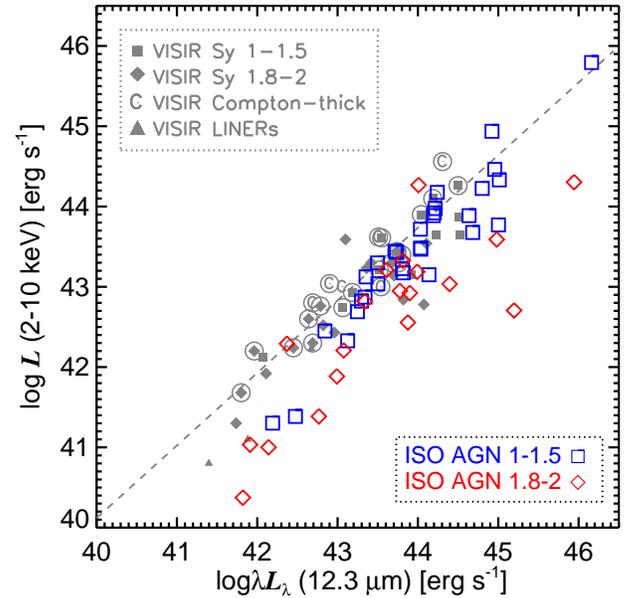}
    \caption{Our correlation from Fig.~\ref{fig:correlation}, compared with the distribution of sources from \citet[][ cf. their Fig.~5]{lutz04}. Their 6 \micron\ continuum luminosities have been converted to 12.3 \micron\ using median $k$-corrections from \citet{hao07}. Luminous unobscured AGN (1--1.5; blue empty squares; those least biased by contamination) fall smoothly onto our correlation in the quasar regime.
    \label{fig:mirxlutz}}
  \end{center}
\end{figure}

We have tested this on the sample of \cite{lutz04}, for which the authors have carefully decomposed AGN and host galaxy contributions using \iso\ spectra. The sample includes Seyferts, quasars as well as ultra-luminous infrared galaxies (ULIRGs), and extends out to $z\approx 1$. We converted the 6 \micron\ continuum luminosities of the 49 sources used in Fig.~\ref{fig:hists} to 12.3 \micron\ using the $k$-corrections of \cite{hao07} for Sy 1s and Sy 2s below \lx=10$^{44}$ erg s$^{-1}$, and using the corresponding quasar SEDs above this limit; but we note that the conclusions below do not depend upon these corrections. The result is shown in Fig.~\ref{fig:mirxlutz}. 

It is seen that the \iso\ sample deviates significantly from our best-fit correlation at low luminosities but not at high luminosities, just as might be expected due to contamination at low powers. The match is particularly good for the brightest unobscured quasars (with log\lx$\ga$44) whose observed emission should be the least biased by any extended SF. 

The deviations of the type 2 AGN, on the other hand, are large even for luminous objects. The sense of the deviation is such that a higher IR power is observed at any given \lx, compared to the distribution of type 1 AGN. Examining the three most luminous type 2 AGN, we find that all of them are actually ULIRGs: IRAS~23060+0505 \citep{brandt97}, IRAS~05189--2524 \citep{severgnini01} and Mrk~463 \citep{bianchi08}. 
The observed core structure of ULIRGs is quite different from Seyferts and quasars. Many ULIRGs are seen to occur in strongly interacting and merging systems where gas gets channeled to the cores and drives very compact and powerful SBs. Two X-ray emitting AGN are sometimes seen in close proximity -- one from each of the merging systems. Therefore, no apriori match of ULIRGs (i.e., without extensive corrections) with our Sy mid-IR:X-ray correlation is expected. Furthermore, the X-ray luminosity of at least one of these -- Mrk~463 -- has been revised upwards by new \c\ imaging (see \citealt{bianchi08}, who find a binary AGN with higher \lx\ than previous estimates), thus actually moving it closer to our correlation line. 

With all these caveats in mind, we regard the match of the luminous sources from \citet[][ at least the unobscured ones]{lutz04} to be fair; consequently, we suggest that Eq.~\ref{eq:correlation} can be extended into the quasar regime as well. Interestingly, our correlation is well-straddled by the two correlation slopes quoted by \citet[][ see their Eq.~1]{fiore08_cosmos} for Sy and quasars respectively, in the CDF-S and C-COSMOS fields. Of course, further detailed validation with enlarged samples will be important.

\section{Systematic effects}
\label{sec:systematics}

\subsection{To be or not to be well-resolved}
\label{sec:scale_systematics}

As described in \S~\ref{sec:results}, our threshold for classification of sources as being well-resolved, or not, is based on locating a discontinuity in the \lmir/\lx--vs.--$r_0$/\rsub\ (luminosity ratio vs. normalized scale) parameter space. The reasoning is that for sources at the higher end of the distribution of VISIR core physical size scales (as compared to the torus), 
significant unresolved contamination is allowed and can lead to an increase in the scatter as well as the mean \lmir\ at any given X-ray power; these sources are the \lq less-resolved\rq\ ones; the rest being \lq well-resolved\rq. Due to the empirical nature of our search, though, it has important limitations which we discuss in this section. But we emphasize that the correlation for all 42 sources remains robust (and with a small scatter relative to previous studies) even if no selection based on resolved scale is applied (see Fig.~\ref{fig:hists}).

\begin{figure}
  \begin{center}
    \includegraphics[width=6.5cm,angle=90]{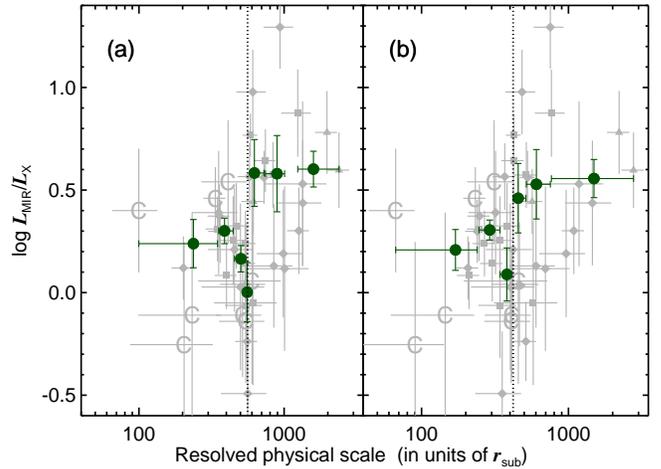}
    \caption{{\em Left (a):} Logarithmic luminosity-ratio vs. normalized resolved scale ($r_0$/\rsub), with \rsub\ being determined according to a constant linear scaling of \lbol--vs.-\lx. The primary adopted threshold of $r_0$/\rsub=560 is shown as the dotted vertical line. Sources falling on the left of this line are taken to be \lq well-resolved\rq, and the rest \lq less-resolved\rq. All 42 sources with symbols as in Fig.~\ref{fig:correlation} are shown in grey, with the errors on scale accounting for uncertainty in \lx\ and slight variations in mid-IR seeing. Bins in $r_0$/\rsub\ are shown by the dark-green bold circles with horizontal error-bars denoting full bin-widths. The standard deviation of the mean log-luminosity ratio in each bin is shown by the vertical error-bars on these circles. {\em Right (b):} In this plot, \lbol\ (and hence \rsub) is determined according to the bolometric corrections of \citet{marconi04}. The threshold is the best-fit discriminatory scale at $r_0$/\rsub=420. See \S~\ref{sec:scale_systematics}. 
    \label{fig:mirxscale_marconi}}
  \end{center}
\end{figure}

Firstly, adding the targets from this paper to the luminosity-ratio vs. normalized scale plot of Paper II (Fig.~3 of that work), the full new distribution is plotted in the left-hand part of Fig.~\ref{fig:mirxscale_marconi}. A crude uncertainty in the normalized scale threshold can be estimated simply from the mean errors of the binned source distribution around the adopted threshold of $r_0$/\rsub=560, which suggests $\Delta (r_0$/\rsub$)\approx 40$. 
Thus increasing (decreasing) our threshold to $r_0$/\rsub=600 (520), the number of well-resolved sources changes from 22 to 25 (16). The corresponding best-fit slope steepens (flattens) to 1.21 (1.07). The details of these fits are listed in Table~\ref{tab:correlationfits} as \lq Well-resolved (Systematic 1$\uparrow$)\rq\ and \lq (Systematic 1$\downarrow$)\rq\ respectively. 

Secondly, in Paper II, a linear scaling of \lbol=10\lx\ was assumed in the computation of \rsub. On the other hand, Bolometric corrections may increase with source luminosity (as discussed in the previous section), 
in which case \rsub\ will increase faster than implied by the linear dependence above, and will result in more of the luminous sources satisfying any given threshold in $r_0$/\rsub. Using the \fbol--\lx\ relation of \cite{marconi04}, we have recomputed \rsub\ (and hence the normalized scale $r_0/$\rsub) as a function of log(\lmir/\lx); the resultant distribution is shown as the light grey points in the right-hand panel of Fig.~\ref{fig:mirxscale_marconi}. On average, the normalized scales are shifted to lower values due to the faster increase in \rsub. In order to best determine the scale where the average distribution of log(\lmir/\lx) undergoes a change, we adopted a floating threshold in normalized scale. The distributions above and below this floating threshold were compared with a two-sided KS test. A value of $r_0/$\rsub=420 proved to be the best discriminator, with the two distributions (well-resolved vs. less-resolved) differing at 98.6 per cent. This can also be roughly discerned by eye in the plot as the noticeable change in the binned distribution; the mean and scatter of the resultant well-resolved and less-resolved sub-samples are ($\bar{r}$, $\sigma_r$)=(0.16, 0.27) and (0.45, 0.38) respectively. The number of sources which satisfy this well-resolved threshold is 22 (a slightly different set of 22 from that discussed in all previous sections) and the resultant \lmir--\lx\ correlation fit is largely unchanged
; see details in Table~\ref{tab:correlationfits} under heading \lq Well-resolved (Systematic 2)\rq. 
So the correlation parameters are insensitive to changes in the resolution threshold, as long as the corresponding well-resolved sub-sample is self-consistently defined.

Lastly, we note that while recent work has confirmed the dependence of \rsub$\propto$$\sqrt{L}$ used in Eq.~\ref{eq:rsub} \citep{suganuma06}, the absolute normalizing scale (in pc) of the relationship remains uncertain. \citet{kishimoto07} have shown that the innermost torus radii inferred from near-IR reverberation mapping of type 1 AGN are systematically smaller by a factor of $\sim$3 than would be implied by the widely-used canonical assumptions implicit in Eq.~\ref{eq:rsub}. They discuss various alternatives to account for this, including a sublimation temperature higher than that used for graphite ($\sim$1500 K), or a typical grain size of $\sim$2 \micron\ -- about four times larger than usually quoted. In any case, a smaller \rsub\ will result in a simple translation of all our sources to larger x-axis ($r_0$/\rsub) values in Fig.~\ref{fig:mirxscale_marconi}. The large change in log(\lmir/\lx) will then be seen to occur around $1700 \times r_0/r_{\rm sub}$ (using the factor $\sim$3 offset found by \citeauthor{kishimoto07}). Again, if the threshold is consistently chosen, there is no effect on our correlation.

\subsection{Treatment of errors}
\label{sec:errors}

The individual X-ray error bars are a significant source of uncertainty for the correlation. Our strategy has been to preferentially choose published analyses with broad-band spectral modeling, and to assign an error based on the range of source variability and fit parameters inferred in the recent literature. If the IR is indeed due to reprocessing in the torus, then the relevant quantities to compare are not necessarily X-ray observations obtained {\em simultaneously} with the IR, but rather the long-term (few years; relevant for a pc-scale torus) X-ray average flux. The IR flux, on the other hand, is expected to be much less variable. The range of published fits spanning various missions and several years (in addition to spectral complexities) results in X-ray error bars much larger than the corresponding IR ones. 

For simplicity, we adopted equal-length (high and low) error bars in log-luminosity space, which should be approximately correct given the many sources of uncertainty. On the other hand, it may be more natural to use equal-length error bars in linear space because flux (and count) measurement errors are usually normally distributed. Thus directly fitting a power-law of the form \lmir=$a$$L_{\rm X}^{b}$ to the base sample of 22 well-resolved sources and accounting for the equal-length X-ray errors in luminosity, still results in a slope $b=1.11\pm0.10$, i.e. very similar to that of Eq.~\ref{eq:correlation}; see Table~\ref{tab:correlationfits} \lq Systematic 3\rq. 

Finally, one may question the veracity of the statistical nature of our errors and ask what happens if we ignore these completely. We have tried a fit according to the algorithm of \citet{isobe90} which can be used when the error assignment and the nature of scatter about a linear relation are ill-understood. We directly used the log-luminosities from the last two columns of Table~\ref{tab:fluxes} (and corresponding ones from Paper II) without the tabulated errors. Adopting the Ordinary Least-square Bisector estimate recommended by \citeauthor{isobe90} results in a regression slope of $b=1.02 \pm 0.07$, i.e. flatter than previous values and consistent with unity at 1-$\sigma$ confidence. This procedure effectively gives identical weighting to all sources, including highly obscured ones; but it is precisely these last sources that tend to have larger inherent uncertainties (e.g. when correcting for Compton-thick obscuration) and that tend to lie above the correlation line in Fig.~\ref{fig:correlation} (see discussion that follows in \S~\ref{sec:irextinction} and \S~\ref{sec:comptonthick}). We thus caution that such a fit may not be wholly appropriate, though we include it for completeness. The fitted slope and intercept are listed in Table~\ref{tab:correlationfits} under \lq Systematic 4\rq.

\subsection{Filter selection}
\label{sec:filter_systematics}
Depending on the redshift of each source, we have tried to choose narrow filters with central wavelengths that ought to be dominated by continuum emission. But we cannot rule out contamination due to emission line wings in every case. For instance, using the median Sy~2 SED from \citet{hao07}, we estimate a worst-case contamination of the \neii~\l12.81 emission line in our NEII\_2 filter photometry of about 14 per cent for the observation of NGC~3281. 

As mentioned, we observed most sources in a number of filters around 12-\micron. We can, therefore, gauge any overall bias on the correlation fit due to selection of our particular filters by choosing instead the second most-appropriate filter from our set, i.e. the one whose central wavelength is closest to 12.3-\micron, following the primary filter. These filters, and the related fluxes for the new targets in this paper, are listed in Table~\ref{tab:fluxes} as \lq secondary\rq\ filters. Including data from Papers I and II, we had access to 34 secondary filters fluxes out of the 42 sources; for the rest, we retain the primary fluxes. Converting these fluxes to 12.3-\micron\ luminosities as before, and fitting the new \lmir--\lx\ correlation reassuringly results in a negligible change of the correlation fit parameters. Details are listed in Table~\ref{tab:correlationfits} under the rows headed \lq Systematic 5\rq. The dispersion in the correlation ($\sigma_r$), on the other hand, does exhibit an increase. This is probably a consequence of the extra contamination in some of the bluer and broader secondary filters (i.e. SIV, SIC, PAH1; with central wavelengths around $\sim$9--12 \micron) due to Silicate absorption, PAH and other features. 
\vspace*{0.2cm}

\noindent
In summary, the considerations of the three preceding sub-sections suggest systematic variations in the mid-IR:X-ray correlation slope over $\sim 1.02-1.21$, in addition to statistical fit uncertainties.

\subsection{Infrared extinction corrections?}
\label{sec:irextinction}

No extinction corrections have been applied to the observed infrared fluxes in any of the above analysis. If, indeed, the tori in these sources are better described by high factors of clumpiness \citep[e.g., ][]{hoenig06} with discrete co-spatial clouds being responsible for both emission and obscuration, then such corrections will be invalid. As such, our tight correlation in Fig.~\ref{fig:correlation} using {\em uncorrected} IR fluxes seems to be an aposteriori fact supporting (but not proving) the non-applicability of extinction corrections. Some previous studies have applied extinction corrections \citep{krabbe01}, while others have not \citep{lutz04}. In this section, we comment on the extent that canonical corrections would have on the inferred luminosities and the \lir:\lx\ correlation.

For our entire sample, the median ${\rm log}N_{\rm H}\sim 23.1$. Using the standard interstellar extinction law \citep{riekelebofsky85}, the extinction $A_{\rm 12 \mu m}$=0.03\av. A typical Galactic gas:dust ratio \citep{bohlin78} then implies a median increase of the \lmir\ values by a factor of 7 for our sample, if one were to correct the mid-IR fluxes for such reddening. On the other hand, for sources with higher obscuration, this correction will far over-estimate the intrinsic power. For instance, the correction factor for a source obscured by a gas column of \nh=4.5$\times$$10^{23}$ cm$^{-2}$ (the median value of our obscured sample) is about 1000; a corresponding increase in \lmir\ would end up pushing the Bolometric luminosities well above the expected Eddington luminosities.

A possible solution may be to use smaller dust:gas ratios. \av/\nh\ values lower by a factor of $\sim 10$ than the Galactic value have been inferred for nearby Seyferts \citep{maiolino01}; this would moderate some of these large corrections, but would still over-estimate the corrections for the extreme Compton-thick cases. We note that most of the sources with the lowest \lmir/\lx\ ratios (the ones highest above the correlation line in Fig.~\ref{fig:correlation}) are all highly obscured and one of these (NGC~7674) is Compton-thick. Recently, Zakamska et al. (2008) have found that deep Si absorption is more likely to occur in AGN with the highest obscuration only. The median Si optical depth for their likely Compton-thick sources is $\tau_{9.7}\approx 1.5$, but with a significant spread of $\tau_{9.7}\sim 1$ in their small sample (similar spreads have been inferred by \citealt{shi06} and \citealt{hao07}). Assuming our highly-obscured sources also have similar IR opacities, and using a standard Milky Way opacity curve \citep{draine03} between 9.7 and 12-\micron, correction factors of $\Delta {\rm log}$\lmir=0.1--0.4 (corresponding to \av=7--35) should be applied to obtain the intrinsic 12.3-\micron\ power of Compton-thick sources. Adopting such corrections has only a small effect on the fitted correlation parameters in Table~\ref{tab:correlationfits}, changing them by less than their 1-$\sigma$ uncertainties.

To reiterate, we do not apply any IR extinction correction.

\section{Compton-thick sources}
\label{sec:comptonthick}

Reliably identifying and gauging the intrinsic powers of Compton-thick (CT) AGN remains a difficult issue irrespective of source redshift \citep[e.g. ][]{comastri07, nandraiwasawa07, alexander08_reliable}. In most sources with extreme obscuring columns, indirect probes have to be employed and allowances made for extinction and contamination on extended scales around the nucleus (where the indirect flux emerges from). It may well be the case that {\em both} the X-ray and IR fluxes require additional corrections for obscuration and dust extinction. So there is no apriori reason to expect such sources to fit the correlation. 

It is therefore encouraging, if somewhat surprising, that our eight CT AGN with \lognh$\ge$24.2 do not deviate significantly from the best-fit correlation. Due to the typically large uncertainties in their luminosities (see details in Appendix of this paper and of Paper II), CT AGN do not have a substantial effect on the correlation fit. But even ignoring these errors, the absolute scatter in log(\lmir/\lx) is still moderate (see Fig.~\ref{fig:hists}).

The importance of our study lies in the fact that the mid-IR:X-ray correlation provides not only an independent, but also accurate (with small scatter) way to probe AGN, irrespective of any obscuration. Simple, high-resolution mid-IR imaging now provides a new proxy for AGN power including CT AGN. We can test this by searching for high-resolution mid-IR imaging observations of other CT Seyferts. Recently, \citet{hoenig08_baldwin} obtained high spatial-resolution $N$-band spectroscopy with VISIR of a small sample of local AGN and discovered a strong mid-IR emission line Baldwin effect. Included in their sample are three CT AGN, of which NGC~1068 is already part of our sample. The second object, Circinus, is a special case (as discussed by these authors) and also showed significant extended emission; therefore, we do not consider it here. This leaves ESO~428--G014, which was first suggested as being CT from \sax\ observations by \citet{maiolino98}. 
\citet{hoenig08_baldwin} quote a 12-\micron\ flux of 0.22~Jy $=>$ log\lmir=42.51 at a distance of 22.3~Mpc. Our mid-IR:X-ray correlation from Eq.~\ref{eq:correlation} then implies log\lx=42.39\p0.06 -- the error being a 1-$\sigma$ Monte Carlo uncertainty. Comparing with recent determinations of the intrinsic X-ray power based on the Fe K$\alpha$ and the optical \oiii~\l5007 emission lines, \citet[][]{levenson06} estimate log\lx=42.40--42.81 for ESO~428--G014 (after applying a small distance correction for consistency). This overlaps with the predicted log\lx\ from our correlation to within 1-$\sigma$. 

The mid-IR and X-ray observations for ESO~438--G014 are over-plotted in Fig.~\ref{fig:correlation} and provide additional validation of our correlation in the CT regime. Of course, this result is based on a relatively-small sample of eight CT sources only, and 
further observations of an enlarged sample will be important to make these conclusions definitive.

There is one final noteworthy observation regarding star-formation around CT AGN. If nuclear starbursts occur in disks aligned with the obscuring AGN tori, they should lie along our line-of-sight to the core and contaminate the nuclear flux, irrespective of angular-resolution; in this case a systematically higher \lir/\lx\ is expected in our CT point-like cores, which we do not observe. This seems to suggest the absence of {\em any} on-going star-formation (even out to several kpc) around local CT sources at the present epoch, at least to within the power limits that we discuss in \S~\ref{sec:starbursts}.

\section{Summary}
\label{sec:summary}

In this work, we suggest that high-resolution mid-infrared continuum photometry is an accurate proxy for the intrinsic X-ray luminosity of local Seyferts. The emitted powers at $\sim$12-\micron\ and 2--10 keV are intimately correlated over three orders of magnitude in luminosity, and evidence suggests that this extends into the quasar regime. Interestingly, it has been noted in a preprint by \citet[][ based on high-resolution Keck imaging]{grossan04} that such a mid-IR:X-ray correlation may be valid for low-luminosity AGN as well. The suitability of the 10--12 \micron\ band for AGN selection has been emphasized by other earlier studies as well \citep[e.g. ][]{spinoglio89, alonsoherrero01}. Our tight correlation now extends this to the highest spatial resolution available, for Seyferts of all types including very heavily obscured sources, and places the conclusion on a precise, quantitative footing with very small scatter.

Combining our sample with Papers I and II, we have presented the most unbiased mid-IR fluxes to date of 41 AGN (not including NGC~1068, which was published by others) -- 16 of these in the current paper. Selecting a sub-sample of sources based on a physically-motivated (but empirically-calibrated) threshold of physical scale resolution related to the torus size, we find the correlation to be especially tight: \lmir$\propto$\lx$^{1.11\pm 0.07}$ (Eq.~\ref{eq:correlation}) with additional systematic variation over the range of 1.02--1.21. We also find a well-defined range in mid-IR:X-ray spectral indices (\alphaix=--1.10\p0.01) for this sample, and show that mid-IR bolometric corrections (\lbol/\lmir, indirectly derived from X-rays) range over $\sim 10-30$ for the Sy luminosity range that we probe.

It is important to reiterate that even without any selection based on resolved scale, our full dataset still has a well-defined distribution in \lmir--vs.--\lx\ (Fig.~\ref{fig:hists}). In fact, if one has access to mid-IR imaging of distant AGN (in effect, low-resolution imaging) and wishes to use this to determine intrinsic X-ray powers, then using the parameters for the correlation fit to all 42 sources (see \lq All\rq\ row in Table~\ref{tab:correlationfits}) is probably more appropriate. Similarly, if one wishes to predict total mid-IR fluxes for small-telescope imaging based on known \lx\ values, the same parameters ought to be used. On the other hand, if the goal is a determination of the intrinsic mid-IR and X-ray emission of the AGN+torus, then the parameters for the well-resolved sample (i.e. Eq.~\ref{eq:correlation}) can be used directly.

There are several other implications of our work, and resultant questions, that we have touched upon only briefly in Papers II and herein. That mid-IR nuclear starbursts cannot be energetically-dominant within the central $\sim 70$ pc of local Seyfert nuclei was discussed in \S~\ref{sec:starbursts}. Our tight dispersion implies that SBs account for no more than about 40\% of the mid-IR emission that remains unresolved with the VLT, on average; this is to be considered as a strong upper-limit, since other effects that increase the dispersion must also be present. The \herschel\footnote{http://www.esa.int/science/herschel} satellite \citep[][ scheduled for launch in 2009]{herschel}, with its large 3.5-m diameter mirror and far-IR detectors, will provide an excellent opportunity to test this result with sensitive (spatial and spectral) deconvolution of SBs from AGN, better than any other space observatory to date. On a more immediate timescale, the Infrared Camera (IRC) on Japan's \akari\footnote{http://www.ir.isas.jaxa.jp/ASTRO-F/} IR satellite has several filters sensitive around the 12-\micron\ regime, one of which (S9W) has been used for an all-sky survey more sensitive than \iras\ by an order of magnitude \citep{akari, akari_irc, akari_allsky}. Our even-higher-resolution 8-m photometry and Seyfert correlation will give a very accurate scaling for removal of contamination in broad-band spectral modeling of AGN with these and other future missions. 

In this regard, we have not speculated on what the resolved scale threshold of $\sim$400--1700 \rsub\ (\S~\ref{sec:scale_systematics}) could correspond to physically. If this is indeed $\sim$10 times larger than typical outer torus radii found in recent models \citep{hoenig06}, then it may instead correspond to a transition between inner and outer nuclear star-forming components \citep[cf. ][ who found a inner star-formation scale of $\sim 50$ pc]{davies07}. But more work is needed to verify this threshold and understand its origin. 

Next, the increase of bolometric corrections -- in other words, a decrease of \lmir/\lbol\ -- with increasing \lbol\ is qualitatively similar to that inferred in the quasar regime by other authors \citep[e.g. ][]{maiolino07, treister08} and may be interpreted in terms of a decrease of covering factor of {\em dust} for more powerful sources. Our correlation extends this result at least one order of magnitude into the Seyfert regime. Comparison with the steeper slope of the X-ray bolometric corrections with luminosity (Fig.~\ref{fig:bolcorr}) has consequences for the covering factor of obscuring {\em gas}. For instance, the nearly-linear luminosity correlation may imply that the gas as the dust are closely coupled together at all powers. But we have not explored this in depth.

\begin{figure*}
  \begin{center}
    \includegraphics[width=8cm,angle=90]{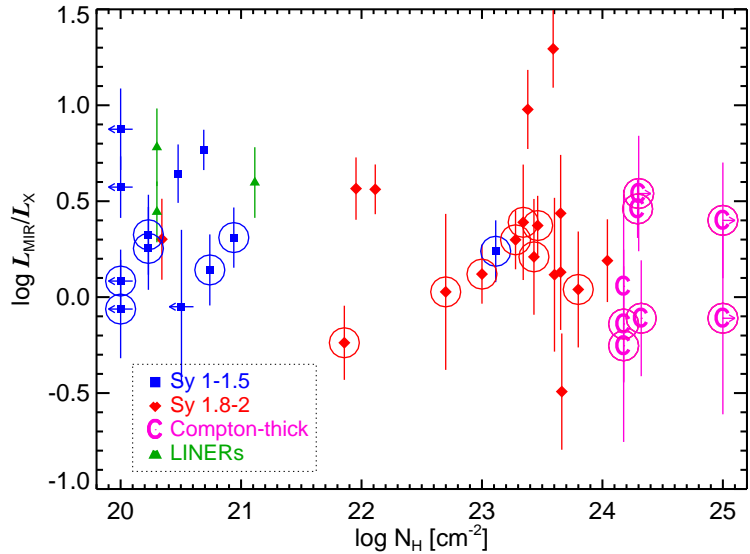}
    \caption{Log luminosity-ratio as a function of obscuring column density determined from X-ray observations (for all sources excluding ESO~323--G032). ($\bar{r}$, $\sigma_r$) = (0.24, 0.38) and (0.37, 0.32) for the 24 obscured (\lognh$\ge$22) and 17 unobscured sources respectively. The corresponding values for the well-resolved sample are (0.17, 0.25) and (0.12, 0.21) with 15 and 7 sources respectively, with the two classes being indistinguishable according to a KS test. \nh\ uncertainties are not included in the plot, but this is irrelevant from the perspective of the absence of any trends. See \S~\ref{sec:summary}.
    \label{fig:mirxnh}}
  \end{center}
\end{figure*}

Finally, the fact that all types of Seyferts and Compton-thick AGN follow the same correlation suggests a similar origin of the mid-IR radiation in all these AGN classes. A non-thermal component is probably not dominant in most cases, as discussed in \S~\ref{sec:corecomposition}. It may be best explained in the context of clumpy AGN tori discussed in Paper II (and references therein). This can also be emphasized by plotting the ratio of mid-IR:X-ray luminosities against the obscuring columns \nh: See Fig.~\ref{fig:mirxnh} which is an update of Fig.~2 in Paper II. 
Any difference in the distribution of log(\lmir/\lx) values for obscured sources with \lognh$\ge$22, as compared to less obscured ones, is significant only at 78 per cent according to a KS test. 
Furthermore, the corresponding distributions for the well-resolved sub-samples (obscured vs. unobscured) are completely indistinguishable. If the gas column density is a measure of the torus inclination angle to the line-of-sight, then smooth-dust torus models would predict a lower log(\lmir/\lx) at high \nh\ values, which we do not see. 

These issues must be accounted for when simulating radiative transfer through AGN tori. We leave detailed study to future work including modeling and observations of enlarged and more complete samples.

\begin{acknowledgements}
It is a pleasure to acknowledge the excellent support work carried out by the staff at ESO that made the service mode observations possible. We thank Eric Pantin for making available his VISIR pipeline to some of us. PG thanks Marco Ajello for discussion on his data of ESO~263--G013. He also acknowledges David R. Ballantyne, David M. Alexander, Richard F. Mushotzky and the anonymous referee for comments and criticisms.

PG was supported by JSPS and RIKEN Foreign Postdoctoral Researcher fellowships during the course of this work. AC, RG and CV thank for partial support the Italian Space Agency (contracts 
ASI--INAF I/023/05/0 and ASI I/088/06/0) and PRIN--MIUR (grant 2006-02-5203).

This research has made use of: 1) the NASA/IPAC Extragalactic Database (NED) which is operated by the Jet Propulsion Laboratory, California Institute of Technology, under contract with the National Aeronautics and Space Administration; 2) the TARTARUS database, which is supported by Jane Turner and Kirpal Nandra under NASA grants NAG5-7385 and NAG5-7067; 3) the Multimission Archive at the Space Telescope Science Institute (MAST). STScI is operated by the Association of Universities for Research in Astronomy, Inc., under NASA contract NAS5-26555. 
\end{acknowledgements}

\bibliographystyle{aa}
\bibliography{gandhi09_mirxray}


\Online

\begin{appendix} 
\section{Notes on individual sources}

\subsection{Compton-thin sources}

\subsubsection{\swiftja\ and \swiftj0601}

The Sy type classification for \object{\swiftja} comes from \citet[][ and references therein]{bodaghee07}; that for \object{\swiftj0601} from \citet{landi07}. The two sources were only recently identified as highly obscured AGN with very small scattering fractions by \citet{ueda07}. Theirs is the best broad-band analysis (combining \suzaku\ and \swift\ data), and we use their luminosities directly after correction for transmission efficiency. Some variability (by factors of 1.6 and 0.5 respectively) as compared to the results of \citet{tueller08} was discussed by \citet{ueda07}; we thus assign factor of 2 luminosity uncertainties. 

\swiftj0601 was the least significant of all our VISIR targets, barely detectable at the 5-$\sigma$ level. A single source, offset by $\Delta$(Right Ascension)=--1\farcs 6, $\Delta$(Declination)=+2\farcs 3 from the nominal position taken from the NASA Extragalactic Database, was found in several filters. Other VLT optical acquisition images obtained by us as part of a separate program confirm that this source coincides with the peak of the optical emission from this highly-reddened, edge-on galaxy.

Both targets were interpreted by \citet{ueda07} as being buried within geometrically-thick tori, and it is interesting to note that one of them (\swiftja) lies {\em above} the best-fit correlation line in Fig.~\ref{fig:correlation} (i.e. is comparatively-weak at 12-\micron, for its X-ray luminosity), possibly implying that hot dust clouds are smothered within the thick tori and not observed directly. Sensitive observations over a broad wavelength range will be required to confirm the nature of this source.

\subsubsection{\object{ESO~209--G12}}

\citet{sazonov07a} state $F_{\rm 17-60\ keV}=1.66\times 10^{-11}$ erg s$^{-1}$ cm$^{-2} =>$ log$L_{17-60}=43.75$ and \lognh$<$22 based on their \integral\ and \rosat\ analysis. Using \xspec\ \citep{xspec} and assuming an unabsorbed power-law (PL) model with $\Gamma=1.7$ and the above normalization also agrees well with $L_{20-100}$ stated by \citet{bassani06} when extrapolated to higher energies. This PL then implies $L_{2-10}=4.45\times 10^{43}$ erg s$^{-1}$. The model straddles the 20--40 keV luminosities stated by \citet{bird06} and \citet{bird07}, thus giving a simple uncertainty estimate between these.

\subsubsection{\object{NGC~3081}}

\citet{beckmann07} find, from a combined \swift\ and \integral\ analysis, $\Gamma =1.8\pm0.2$ with $L_{20-100}=8.1\times 10^{42}$ erg s$^{-1}$. They also find significant variability of a factor of a few in the BAT 14--195 keV lightcurve over a 15 month period.

Fitting an absorbed PL above a few keV (the softer flux is fit by weak scattering or a blackbody) to spectra extracted from the Tartarus database gives \nh=$68(\pm 5) \times 10^{22}$ cm$^{-2}$ with $L_{2-10}=4-6\times 10^{42}$ erg s$^{-1}$; the variation being that between the \asca\ SISs. Extending the de-absorbed PL to 200 keV gives $L_{14-195}=9-14\times 10^{42}$ erg s$^{-1}$, well matching the more recent values quoted by \citet{markwardt05} and \citet{ajello08_swift}. This suggests that variability may not be very strong. 

On the other hand, \sax\ is known to have observed the source in a low flux state \citep{maiolino98}. We take the lower value of the luminosity range found in the above \asca\ fits, with a large error of \logl=0.3 as a mean value.

This source was also imaged by \citet{krabbe01} at 1\farcs3 resolution. The full $N$-band flux quoted by them is consistent with our narrow-band PAH2 flux (assuming a flat SED in $\lambda F_{\lambda}$ units), implying the absence of extended emission on $\sim$ 1\arcsec\ scales.

\subsubsection{\object{ESO~263--G013}}

The only good-quality X-ray spectrum published is the \swift\ XRT analysis by \citet{landi07}, who find \nh=4\p1 $\times 10^{23}$ cm$^{-2}$ absorbing a PL with $\Gamma\approx 2.32$. The observed flux is $F_{2-10}=2\times 10^{-12}$ erg s$^{-1}$ cm$^{-2}$ and they quote an intrinsic $L_{2-10}=4-7\times 10^{42}$ erg s$^{-1}$ (converted to our cosmology). But extending this steep PL to higher energies underpredicts other \integral\ and \swift\ fluxes by factors of $\sim 7-10$ \citep{bird07, ajello06}. Recent \xmm\ data (by M. Ajello, priv. comm., and some of us) analysed in conjunction with the \swift\ BAT data are instead fit with a harder photon-index $\Gamma\approx 1.8$, resulting in reasonable agreement (to within a factor of $\sim 2$) with fluxes over 2--100 keV and an intrinsic $L_{2-10}\sim 2\times 10^{43}$ erg s$^{-1}$; we prefer this latter broad-band analysis.

\subsubsection{NGC~4388}

Combined analysis from several missions shows a simple PL above a few keV absorbed by a column \nh=$2.7\times 10^{23}$ cm$^{-2}$, but with highly variable fluxes and obscuring column \citep{beckmann04, cappi06, iwasawa03, risaliti02_nhvar}. Using the mean of the large range found in \xmm\ by \citet{cappi06} and recent \integral\ results by \citet{beckmann06}, for a distance of 16.7 Mpc, results in a mean log$L_{2-10}=42.24$, to which we attach a large 1-$\sigma$ error in dex of 0.3. 

\subsubsection{\object{ESO~323--G32}}

This source is classified as a Sy~1.9 in the catalogue of \citet{veron01}. The only reliable published X-ray measurements happen to be the \integral\ observations of \citet{bird07}, who find $F_{20-40}=6.81\times 10^{-12}$ and $F_{40-100}=1.51\times 10^{-11}$ erg s$^{-1}$ cm$^{-2}$. In the absence of other information, we parametrize this with a simple PL, $\Gamma\approx 1.2$, which extrapolates to $F_{2-10}=3.7\times 10^{-12}$ erg s$^{-1}$ cm$^{-2}$, or $L_{2-10}=2.0\times 10^{42}$ erg s$^{-1}$. One could also predict the 2--10 keV power based on forbidden optical emission line measurements. Correcting the \oiii~\l5007 emission line flux from \citet{gu02} for a small extinction factor from the same paper, and using the type-1 $<$$F_{2-10}/F_{\rm [OIII]}$$>$ relation from \citet{panessa06}, we get $L_{2-10}\approx 3.3\times 10^{42}$ erg s$^{-1}$.

\subsubsection{\object{NGC~4992}}

The Sy classification of this source comes from \citet{bodaghee07} and references therein. \citet{comastri07} have presented a broad-band \suzaku\ spectral analysis, and infer an extreme reflection component with $R>5$ and a possible relativistically-blurred Fe line, indicative of disc-dominated reflection. This can be naturally explained by relativistic light-bending on to the inner disc, which preferentially enhances reflection and suppresses the incident power-law at infinity \citep{miniutti04}. The fraction of AGN which are described by this light-bending model is still unknown, though increasing numbers of sources are being discovered and the hard X-ray background spectrum can also accommodate a non-negligible fraction of light-bent source in the Universe (see for instance \citealt{miniutti04}, \citealt{g07}, \citealt{miniutti07_iras13197} and references therein).

Although the majority of intrinsic power-law photons may be bent towards the event horizon and never escape the black hole's gravitational sphere of influence, the relevant luminosity in our case is the X-ray flux that is incident on the distant torus clouds, i.e. the (absorption-corrected) power observed at infinity. Using the model of \citet{comastri07} yields $L_{2-10}=1.7\times 10^{43}$ erg s$^{-1}$. If part of the high reflection is due instead to decreased transmission efficiency by heavy obscuring material or due to strong time variability, then this estimate will be a lower limit (cf. \citealt{ueda07}).

\subsubsection{\object{NGC~6300}}

This is one of a sample of \lq changing-look\rq\ AGN whose spectra exhibit transitions between Compton-thick and Compton-thin obscuration on timescales of years \citep{matt03_changinglook}. It was found to be reflection-dominated in 1997, but has since been in a Compton-thin state \citep[\rxte\ and \sax\ analysis; ][]{guainazzi02}. Assuming a Compton-thin column of \nh=$2\times 10^{23}$ cm$^{-2}$ (from the \xmm\ analysis of \citealt{matsumoto04}) obscuring a PL with $\Gamma=2.42$ (\citealt{beckmann06}) and normalizing to JEM-X and ISGRI 2--100 keV fluxes of the latter publication, the implied intrinsic log$L_{2-10}=42.76$. But, in addition to its changing-look nature, the source also exhibits large flux variability \citep{awaki06} and a much lower luminosity of log$L_{2-10}=41.8$ was derived by \citet{matsumoto04} from their \xmm\ analysis. We use a mean log$L_{2-10}=42.3$.

\subsubsection{\object{ESO~103--G35}}

\citet{molina06} analysed the broad-band \integral\ spectrum of this source and located an exponential cut-off to the X-ray PL at $E_{\rm cut}\approx 70$ keV, with no substantial flux change from previous \sax\ observations \citep{risaliti02_bright, wilkes01}. This source has shown moderate and fast \nh\ variability on time scales of several months \citep{risaliti02_nhvar}. Using model realizations in \xspec, we find that the PL model of \citet{molina06} with $\Gamma=1.78$, \nh=$2\times 10^{23}$ cm$^{-1}$ and $F_{2-10}=3\times 10^{-11}$ cm$^{-2}$ is consistent with the higher energy 20--100 keV flux when extrapolated, but slightly underestimates (by $\sim 30$ per cent) the latest 20--40 and 40--100 keV fluxes from \citet{bird07}. On the other hand, the \swift\ BAT 14--195 keV luminosity is over-estimated by a factor of $\approx 2$. We assume the base 2--10 keV luminosity of the \citeauthor{molina06} model, with an 1-$\sigma$ luminosity error of 0.15 in dex.

\subsubsection{\object{NGC~6814}}

This source is an unabsorbed Sy 1.5, with  \lognh$<$20.7 \citep{reynolds97}. It is known to be highly variable, even after accounting for a neighbouring CV which contaminated early observations \citep[e.g. see the results of long-term \rxte\ monitoring described by ][]{mukai03}. Using the PL model with $\Gamma=2.48$ from \citet{beckmann06}, and extrapolating from their \integral\ 20--100 fluxes to below 10 keV suggests that JEM-X should have detected the source, which it did not. Instead, some spectral curvature probably needs to be invoked. Including a simple \pexrav\ \citep{pexrav} parametrization with $\Gamma=1.6$, $R=1$ and $E_{\rm cut}=150$ keV agrees with 20--40 keV and 40--100 keV fluxes of \citet{bird07}, agrees with the mean published \rxte\ observations mentioned above, and also matches the $<10$ keV photon-index found from \asca\ analysis of data in the Tartarus database. This implies a mean $L_{2-10}=1.33\times 10^{42}$ erg s$^{-1}$. A large variation of 0.4 in dex accounts for the source variability.

\subsection{Compton-thick sources}

\subsubsection{NGC~1068}

NGC~1068 is one of the best-studied of obscured AGN; hence our decision to include it on our correlation. In X-rays, it is known to be substantially Compton-thick and reflection-dominated, with \lognh$>$25 \citep{matt00, iwasawa97, koyama89}. Correcting from the observed flux of X-rays scattered from above the obscuring medium into the line-of-sight, or from the flux of the forbidden \oiii~\l~5007\AA\ emission line, intrinsic log$L_{2-10}$ estimates of $\sim 42.8-44$ are derived (\citealt{iwasawa03, panessa06, levenson06}). 

In the mid-IR, several high-resolution studies have been carried out from the ground revealing bright, extended structures on sub-arcsec scales and beyond \citep[][]{galliano05_ngc1068, bock00, alloin00}. At the highest spatial sampling, continuum core fluxes of 9--10 Jy have been reported around 12-\micron\ \citep{mason06, tomono01}; these correspond to log$\lambda L_{\lambda}\approx 43.8$.

\subsubsection{\object{NGC~3281}}

\citet{vignalicomastri02} identified this source as being mildly Compton-thick. Using their model (c) which corrects for Compton-scattering in a physically-motivated way $=>L_{2-10}=1.7\times 10^{43}$ and $L_{20-100}=2.1\times 10^{43}$ erg s$^{-1}$ in our cosmology. Using this model to infer the 20--40 keV and 14--195 keV fluxes in order to compare with the recent studies of \citet{bird07} and \citet{markwardt05} suggests log$L_{2-10}=43.30\pm0.15$.

\citet{krabbe01} detected the presence of extended $N$-band emission beyond their resolution limit of 1\arcsec\ in this source. Consistent with this, our 0\farcs35-resolution flux (in units of $\lambda F_{\lambda}$) is even lower than theirs by $\approx 30$ per cent.

\subsubsection{\object{NGC 3393}}

Corrections based on the Fe line and on the flux of forbidden optical \oiii\ emission yield $L_{2-10}\approx 7.2-17\times 10^{42}$ erg s$^{-1}$ \citep{levenson06}. Combining \xmm\ analysis with previous \sax\ data, \citet{guainazzi05} find a decreased column density (\nh$ \sim 4\times 10^{24}$ cm$^{-2}$) with respect to the complete Compton-thick suppression (\nh$ > 10^{25}$ cm$^{-2}$) inferred by \citet{maiolino98}. Because only a small part of the broad-band X-ray flux is transmitted, correcting the observed continuum to estimate the intrinsic luminosity is highly uncertain given the current data quality on this source. Still, \citet{levenson06} find reasonable agreement when extrapolating the \sax\ high-energy data to their \c\ bandpasses with $\Gamma\approx 2$. \suzaku\ may provide better estimates in the near future.

\subsubsection{\object{IC~3639}}
The intrinsic high-energy flux from this source is likely to be completely depleted due to heavy Compton-thick obscuration with $N_{\rm H}>10^{25}$ cm$^{-2}$ \citep[e.g., ][]{risaliti99}. Analysis is complicated by the fact that most of the $<10$ keV X-ray flux observed is likely related to extended emission \citep{ghosh07, guainazzi05}. Using instead the \oiii~\l5007 emission line ($F_{\rm [OIII]}=3.4\times 10^{-12}$ erg s$^{-1}$ cm$^{-2}$ from \citealt{lumsden01}, corrected for extinction) as a proxy for the intrinsic power (in conjunction with the \oiii:X-ray relation from \citealt{panessa06}) gives $L_{2-10}\sim 4\times 10^{43}$ erg s$^{-1}$. This also matches older \oiii\ measurements quoted by \citet{guainazzi05}, if we correct for extinction using the prescription of \citet{bassani99}.

\subsubsection{\object{NGC~5728}}

From a broad-band \suzaku\ analysis, \citet{comastri07} conclude that this source is Compton-thick with $N_{\rm H}=2.1\times 10^{24}$ cm$^{-2}$. The unabsorbed flux is $F_{2-10}\sim 5\times 10^{-11}$ erg s$^{-1}$ cm$^{-2}$, implying $L_{2-10}\sim 8.7\times 10^{42}$ erg s$^{-1}$. There is additional low-energy spectral complexity due to scattering/reflection, but this accounts for only 1--2 per cent of the 2--10 keV power. The mildly Compton-thick column suppresses the overall X-ray continuum by only $\sim 5-10$ per cent; we include these small corrections.

From the raw \oiii~\l5007 optical emission line measurement of \citet{gu02} along with the type-1 $<$$F_{2-10}/F_{\rm [OIII]}$$>$ relation from \citet{panessa06}, we derive $L_{2-10}\sim 1.1\times 10^{42}$ erg s$^{-1}$. Correcting for extinction using the Balmer decrement yields $L_{2-10}\sim 3.4\times 10^{42}$ erg s$^{-1}$. 

We use the mean of the above luminosities, which also agrees with the extrapolated \swift\ BAT measurement by \citet{markwardt05}, within errors.

\subsubsection{\object{ESO~138--G1}}

\citet{collingebrandt00} found evidence for Compton-thick obscuration in this source, based on their \asca\ spectral analysis and comparison with \oiii\ optical emission line flux, but the reflection fraction and obscuring column were not well constrained. Suppose we assume that the source PL ($\Gamma=2.1$) is mildly Compton-thick with \nh=$2\times 10^{24}$ cm$^{-2}$, reflection $R$=1 with a typical high energy exponential cut-off at 200 keV. Then, normalizing the model to recent \integral\ 20--40 keV fluxes from \citet{bird07} leads to a good match to reported measurements up to 100 keV \citep{bassani06}. In this case, log$L_{2-10}=42.9$ intrinsically, including mild correction for multiple Compton down-scattering \citep{wf}.

On the other hand, using the Sy 1 relationship between the \oiii\ and 2--10 keV fluxes described by \citet{collingebrandt00}, a higher log$L_{2-10}=43.10$ is implied. Given the paucity of good X-ray data on this source, we assume the mean of the above luminosities, with a large uncertainty of 0.3 dex.


\end{appendix}

\end{document}